\documentclass[onecolumn]{emulateapj}

\usepackage{epsfig}
\usepackage{graphicx}
\usepackage{amsmath, amsthm, amssymb}
\usepackage{rotating}
\usepackage{multirow}

\shortauthors{Lim and Tan}

\begin{document}

\title{Far-Infrared Extinction Mapping of Infrared Dark Clouds}

\author{Wanggi Lim}
\affil{Dept. of Astronomy, University of Florida, Gainesville, FL 32611, USA}

\author{Jonathan C. Tan}
\affil{Depts. of Astronomy \& Physics, University of Florida, Gainesville, FL 32611, USA}

\begin{abstract}
Progress in understanding star formation requires detailed
observational constraints on the initial conditions, i.e. dense clumps
and cores in giant molecular clouds that are on the verge of
gravitational instability. Such structures have been studied by their
extinction of Near-Infrared (NIR) and, more recently, Mid-Infrared
(MIR) background light. It has been somewhat more of a surprise to
find that there are regions that appear as dark shadows at
Far-Infrared (FIR) wavelengths as long as $\sim100\micron$! Here we
develop analysis methods of FIR images from {\it{Spitzer}}-MIPS and
{\it{Herschel}}-PACS that allow quantitative measurements of cloud
mass surface density, $\Sigma$. The method builds upon that developed
for MIR extinction mapping (MIREX) \citep{butler2012}, in particular
involving a search for independent saturated, i.e. very opaque,
regions that allow measurement of the foreground intensity. We focus
on three massive starless core/clumps in IRDC G028.37+00.07, deriving
mass surface density maps from 3.5 to $70\:\micron$. A by-product of
this analysis is measurement of the spectral energy distribution of
the diffuse foreground emission. The lower opacity at $70\:\micron$
allows us to probe to higher $\Sigma$ values, up to
$\sim1\:{\rm{g\:cm}^{-2}}$ in the densest parts of the
core/clumps. Comparison of the $\Sigma$ maps at different wavelengths
constrains the shape of the MIR-FIR dust opacity law in IRDCs. We find
it is most consistent with the thick ice mantle models of
\citet{ossenkopf1994}.  There is tentative evidence for grain ice
mantle growth as one goes from lower to higher $\Sigma$ regions.
\end{abstract}


\section{Introduction}

The molecular gas clumps that form star clusters are important links
between the large, Galactic-scale and small, individual-star-scale
processes of star formation. Most stars are thought to form from these
structures \citep[e.g.,][]{lada2003,gutermuth2009}.
Massive stars may form from massive starless cores buried in such
clumps \citep[e.g.,][]{mckee2003}. 

Observations of clumps in their earliest stages of star formation are
thus important for constraining theoretical models of massive star and
star cluster formation. These early-stage clumps are expected to be
cold and dense, and large populations have been revealed as ``Infrared
Dark Clouds'' (IRDCs) via their absorption of the diffuse MIR 
($\sim10\:\micron$) emission of the Galactic interstellar medium (e.g. 
Egan et al. 1998; Simon et al. 2006 [S06]; Peretto et al. 2009; Butler \& Tan 2009).
With the advent of longer wavelength imaging data from
{\it{Spitzer}}-MIPS \citep{carey2009} and {\it{Herschel}}-PACS
\citep[e.g.,][]{peretto2010,henning2010}, some IRDCs are appear dark
at wavelengths up to $\sim100\:\micron$.

Butler \& Tan (2009, 2012 [BT09, BT12]) developed MIR extinction
(MIREX) mapping with {\it{Spitzer}}-IRAC 8$\micron$ GLIMPSE survey
data \citep{churchwell2009}.  Using this method, which does not
require knowledge of cloud temperature, they derived mass surface
density, $\Sigma$, maps of 10 IRDCs with angular resolution of
2$\arcsec$.
The maps probed $\Sigma$ up to $\sim0.5\:{\rm{g\:cm}}^{-2}$
($A_{V}\sim100\:{\rm{mag}}$), with this limit set by the image noise
level and the adopted opacity of $7.5\:{\rm{cm^2\:g^{-1}}}$ (based on
thin ice mantle dust models; \citet{ossenkopf1994} [OH94]).

However, higher $\Sigma$ clouds have been claimed based on FIR/mm dust
emission \citep[e.g.,][]{battersby2011,ragan2012}, although this
requires also estimating the dust temperature. Some already-formed
star clusters have cores with $\Sigma>1\:{\rm{g\:cm}}^{-2}$
\citep[e.g.,][]{tan2013a}. It is thus important to see if extinction
mapping methods can be developed that can probe to higher $\Sigma$
values. Butler, Tan \& Kainulainen (2013 [BTK]) have attempted this by
using more sensitive (longer exposure) 8~$\micron$ images. Here we
develop methods that probe to higher $\Sigma$ via FIR extinction
mapping using {\it{Spitzer}}-MIPS 24 and {\it {Herschel}}-PACS
70$\micron$ images with 6\arcsec\ angular resolution. At these
wavelengths, IRDCs still appear dark but have smaller optical depths.
Our method also requires us to examine the FIR extinction law and its
possible variations within dense gas, which may be caused by grain
coagulation and ice mantle formation.


\section{The Far-Infrared Extinction (FIREX) Mapping Method}\label{S:methods} 

\subsection{MIR and FIR Imaging Data for IRDC G028.37+00.07}

We utilize {\it Spitzer}-MIPS $24\micron$ images from the MIPSGAL
survey \citep{carey2009} that have 6\arcsec\ resolution and
estimated $1\sigma$ noise level of $0.25\:$MJy/sr. 

We also analyze archival $70\micron$ {\it{Herschel}}-PACS images.
The first type (proposal ID KPGT-okrause-1) were observed with medium
scanning speed and also have $\simeq$6\arcsec\ resolution, but do not
cover a very large area around the IRDC ($\sim16$\arcmin\ in extent).
We estimate a $1\sigma$ noise level of $\sim15\:$MJy/sr, i.e. about a
factor of two better than achieved with fast scanning observations
\citep{traficante2011}. The second type (proposal ID KPOT-smolinar-1)
were observed in the Galactic plane HiGAL survey \citep{molinari2010}
in fast scanning mode with $\sim$9\arcsec\ resolution. These data are
useful for assessing intensity of the Galactic background emission,
which needs to be interpolated from regions surrounding the IRDC.

Both the {\it Herschel} data sets are obtained already processed to
level~2.5 in HIPE, so zero-level offsets need to be applied to have
measurements of absolute values of specific intensities.
We adopted a model spectral energy distribution (SED) of the diffuse
Galactic plane emission (Li \& Draine 2001 [LD01]) from NIR to
FIR. We fit this model to the observed median intensities in a
2$\degr\times$2$\degr$ field centered on the cloud, considering data
at 8$\micron$~({\it{Spitzer}}-IRAC), 24$\micron$~({\it{Spitzer}}-MIPS),
60$\micron$ and 100$\micron$ (both {\it{IRAS}}) and then predicted the
expect intensities in the {\it{Herschel}}-PACS~70\micron\ band. A
single offset value was then applied to each {\it{Herschel}}
dataset (908~MJy/sr and 241~MJy/sr for medium and fast scan data,
respectively). We tested this method against offset corrections
reported by \citet{bernard2010} at $l=30\degr$ based on Planck data, finding
agreement at the $\simeq10\%$ level.

\begin{figure}[!tb]
\plottwo{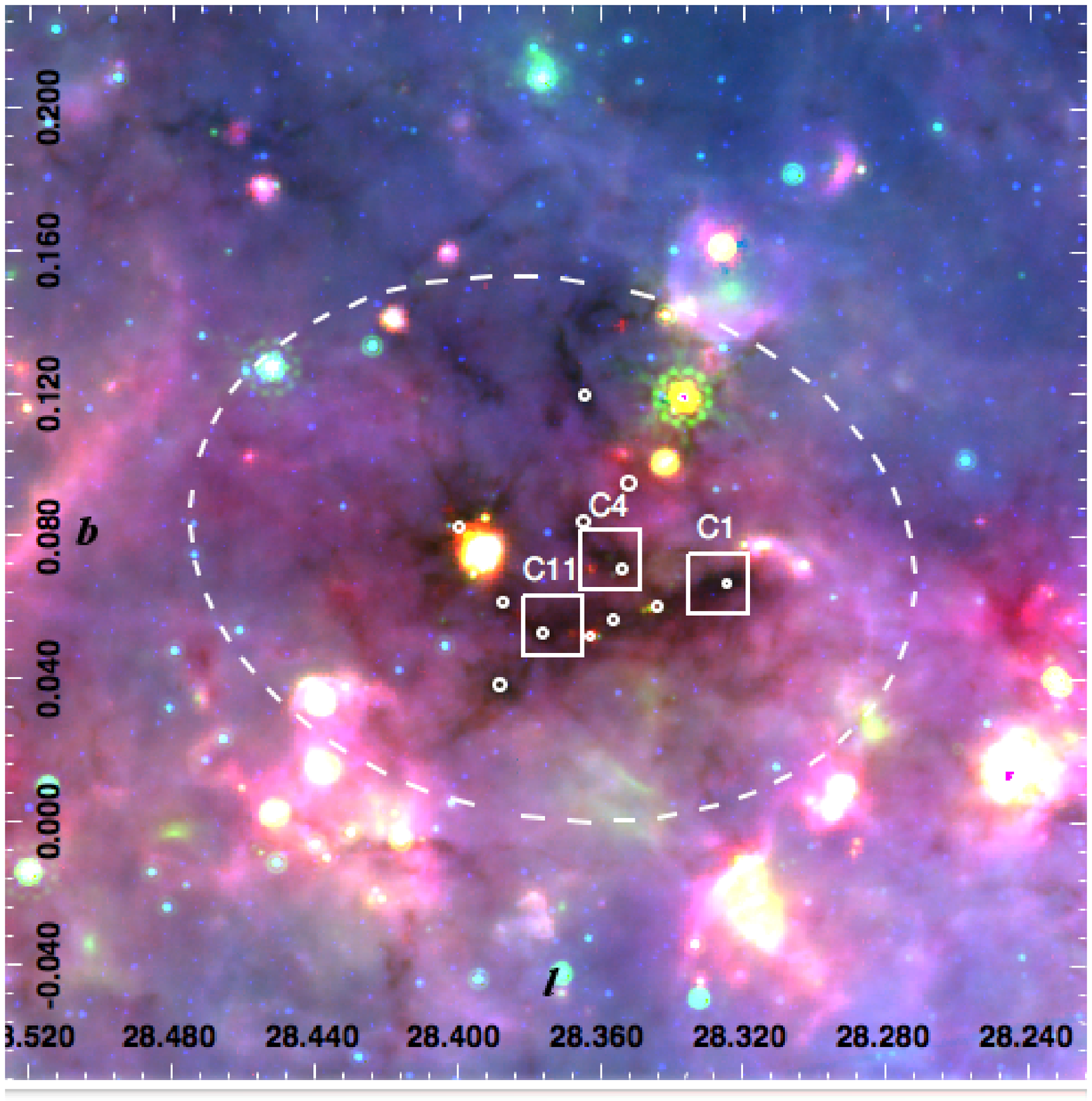}{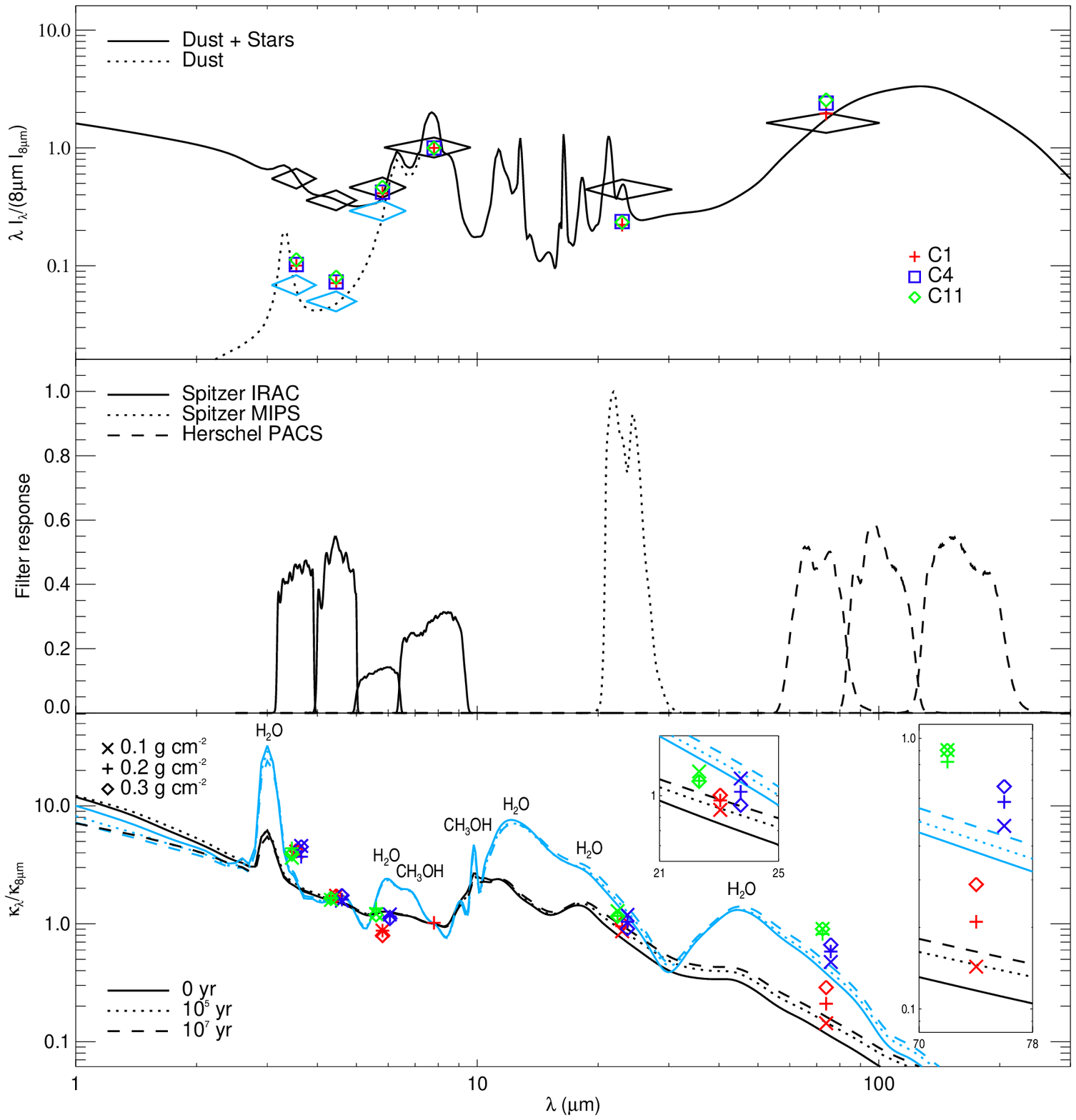}
\caption{
\footnotesize (a) Left: Three-color image of IRDC G028.37+00.07 (red:
70$\micron$, green: 24$\micron$ \& blue: 8$\micron$). Small circles
indicate massive cores studied by BT12 and BTK. Squares are regions
inspected for local saturation around cores C1, C4 \& C11. Dashed
ellipse shows the IRDC boundary from S06.
{\it{(b) Right top:}} Total IR to sub-mm SED (solid line) of
the Galactic plane in the ``MIRS'' region at $l\simeq44\degr$
(LD01). Dotted line shows contribution to this from
dust; stars dominate at short wavelengths.
Large diamonds indicate convolution of these SEDs with the filter
response function of corresponding instruments (panel (b)). The
red/blue/green symbols show foreground intensities, relative to the
LD01 IRAC 8$\micron$ value, measured to the three saturated cores
C1/C4/C11, respectively.
{\it{(c) Right middle:}} Filter response of IRAC bands 1-4 (solid
lines), MIPS~24$\micron$ (dotted line) and PACS~70, 100, \&
160$\micron$ (dashed lines). 
{\it{(d) Right bottom:}} Dust opacities relative to effective opacity
measured in the IRAC~8$\micron$ band. Black/blue lines show thin/thick
ice mantle models of OH94. Solid lines show the initial, uncoagulated
MRN grain population. Dotted and dashed lines indicate these models after
$10^5$ and $10^7\:{\rm{yr}}$ of coagulation at
$n_{\rm{H}}=10^6\:{\rm{cm}}^{-3}$,
respectively. The red/blue/green symbols (for C1/C4/C11 regions;
slight wavelength offsets applied for clarity) show
$\kappa_{\lambda}/\kappa_{8\micron}$ for the indicated $\Sigma$
ranges.  }
\label{fig:wholecloud}
\end{figure}

We find an astrometric difference of a few arcseconds between the
{\it{Herschel}} and {\it{Spitzer}} maps. We corrected this by the
average value of the mean positional offset of four point sources seen
at 8, 24 and 70~$\rm{\mu}m$, amounting to a 3.4\arcsec\ translation at
P.A. 62.7$\degr$. The resulting three-color image of the IRDC is shown
in Fig.~\ref{fig:wholecloud}.

While our focus is on FIR data, we also compare to {\it{Spitzer}}-IRAC~3.5,
 4.5, 5.9, $8\micron$ (GLIMPSE) images, which have
2\arcsec\ resolution at 8\micron. For these, we adopt $1\sigma$ noise
levels of 0.3~MJy/sr, 0.3~MJy/sr, 0.7~MJy/sr, 0.6~MJy/sr, respectively
\citep{reach2006}.

To compare multiwavelength extinction mapping pixel by pixel, we
regrid the IRAC images to the 24\micron\ MIPS frame with its
$1.25\arcsec\times1.25\arcsec$ pixels, similar to the 1.2\arcsec\ IRAC
pixels. However, for {\it{Herschel}}-PACS data, with its
3.2\arcsec\ pixels, we first carry out extinction mapping on the
original pixel grid (especially searching for saturated pixels, see
below), before finally regridding to the MIPS frame for
multiwavelength comparison.



\subsection{Radiative Transfer and Foreground Estimation}

We adopt the 1D radiative transfer model of BT12, which requires
knowing the intensity of radiation directed toward the observer at the
location just behind, $I_{\nu,0}$, and just in front, $I_{\nu,1}$, of
the target IRDC. The infrared emission from the IRDC is assumed to be
negligible (which imposes an upper limit on the dust temperature of
$\sim15\:$K; \citep{tan2013b}), so that
\begin{equation}
I_{\nu,1}=e^{-\tau_{\nu}}I_{\nu,0}
\end{equation}
where optical depth $\tau_{\nu}=\kappa_{\nu}\Sigma$ and $\kappa_{\nu}$
is total opacity at frequency $\nu$ per unit total mass and $\Sigma$
is the total mass surface density. The value of $I_{\nu,0}$ is to be
estimated via a suitable interpolation from the region surrounding the
IRDC, while $I_{\nu,1}$ is to be derived from the observed intensity
to given locations towards the cloud. However, because the observed
Galactic background emission, $I_{\nu,0,{\rm{obs}}}$, and the observed
intensity just in front of the IRDC, $I_{\nu,1,{\rm{obs}}}$, are
contaminated with foreground emission, $I_{\nu,{\rm{fore}}}$, (IRDCs
are typically at several kpc distance in the Galactic plane) we
actually observe
\begin{equation}
I_{\nu,1,{\rm{obs}}}=I_{\nu,{\rm{fore}}}+I_{\nu,1}
\end{equation}
\noindent and 
\begin{equation}
I_{\nu,0,{\rm{obs}}}=I_{\nu,{\rm{fore}}}+I_{\nu,0}.
\end{equation}
Therefore, like MIREX mapping, FIREX mapping requires measurement
of $I_{\nu,{\rm{fore}}}$ towards each region to be mapped.

Following BT12, we estimate $I_{\nu,{\rm{fore}}}$ empirically by
looking for spatially independent dark regions 
that are ``saturated'', i.e. they have the same observed intensity to
within some intensity range set by the noise level of the image. Using
$8\:\micron$ GLIMPSE images with 2\arcsec\ resolution and $1\sigma$
noise level of 0.6~MJy/sr, BT12 defined saturated pixels as being
within a 2$\sigma$ range above the observed global minimum value in a
given IRDC. However, for the IRDC to be said to exhibit saturation,
these pixels needed to be distributed over a region at least
8\arcsec\ in extent.

We follow a similar method here for the images in the IRAC bands at 24
and 70\micron, but with the following differences: (1) We search for
``local saturation'' in smaller 1\arcmin\ by 1\arcmin\ fields of view
that contain dense cores previously identified by BT12. This helps to
minimize the effects of foreground spatial variation. (2) In addition
to a standard $2\sigma$ intensity range we also consider $4\sigma$ and
$8\sigma$ ranges. This is because we regard the estimate of the noise
level in the images as somewhat uncertain. We will gauge the liklihood
of saturation also by considering the morphology (e.g. connectedness)
of the saturated pixels, together with their overlap with saturated
pixels at other wavelengths. In general we expect the saturated region
to be larger at wavelengths with larger dust opacities (i.e. generally
increasing at shorter wavelengths), but the size is also affected by
the relative noise levels in the different wavebands. The ability to
detect saturation can also be compromised by the presence of a nearby
source that enhances the local level of foreground emission. (3) Given
the larger beam sizes at 24 and 70\micron, we adopt a less stringest
spatial extent criterion, requiring separation of saturated pixels by
about two times the beam FWHM, i.e. 12\arcsec.


Our method for estimating the background intensity is the same adopted
by BT09 and BT12, i.e. the small median filter method applied outside
the defined IRDC ellipse from S06 (filter size set to 1/3 of the major
axis, i.e. 4\arcmin), followed by interpolation inside this
ellipse. We inspected the background fluctuations in three control
fields equal to this filter size just outside the ellipse: after
foreground subtraction the average $1\sigma$ fluctuations (from a
fitted Gaussian) were 17.3\%, 25.3\% and 42.3\% at 8, 24, 70\micron,
respectively. In the optically thin limit, these values provide an
estimate of the uncertainty in the derived $\tau$ and thus $\Sigma$ in
any given pixel due to background fluctuations.

\subsection{MIR to FIR Opacities}


Following BT09, we adopt a spectrum of the diffuse Galactic background
from the model of LD01. We will see below that a by-product of our
extinction mapping is a measurement of this spectrum of the foreground
emission.

We then consider several different dust models (Table~\ref{tb:kappa}),
especially the thin and thick ice mantle models for moderately
coagulated (i.e. coagulation for $10^5\:$yr at $n_{\rm{H}=10^6\:{\rm{cm}^{-3}}}$) 
grains (OH94). For all models we adopt a
total (gas plus dust) mass to refractory dust mass ratio of 142
\citep[][c.f. the value of 156 used in BT09 and BT12]{draine2011}.
This choice will not affect measurement of relative opacities.
Finally, we obtain the effective opacities in different wavebands,
i.e. convolution of the background spectrum, filter response function
and opacity function (see Fig.~\ref{fig:wholecloud}b,c,d and Table~\ref{tb:kappa}).


\begin{deluxetable}{lcccccccc}
\tabletypesize{\tiny}
\tablecolumns{9}
\tablewidth{0pt}
\tablecaption{Telescope Band and Background-Weighted Dust Opacities Per Gas Mass\tablenotemark{a} ($\rm{cm^{2}\:g^{-1}}$)}
\tablehead{\colhead{Dust Model\tablenotemark{b}}                                              &
           \colhead{IRAC3.5} &  
           \colhead{IRAC4.5} &  
           \colhead{IRAC6} &  
           \colhead{IRAC8} &  
           \colhead{MIPS24} &
           \colhead{PACS70} &
           \colhead{PACS100} &
           \colhead{PACS160}\\
	   \colhead{} &
	   \colhead{$\rm{3.52\:\micron}$\tablenotemark{c}} &
	   \colhead{$\rm{4.49\:\micron}$} &
	   \colhead{$\rm{5.91\:\micron}$} &
	   \colhead{$\rm{7.80\:\micron}$} &
	   \colhead{$\rm{23.0\:\micron}$} &
	   \colhead{$\rm{74.0\:\micron}$} &
	   \colhead{$\rm{103.6\:\micron}$} &
	   \colhead{$\rm{161.6\:\micron}$} 
}
\startdata
WD01 $R_V=3.1$ & 8.73 & 5.45 & 3.63 & 5.45 & 3.92 & 0.392 & 0.185 & 0.0756\\
WD01 $R_V=5.5$ & 11.5 & 7.59 & 4.86 & 6.13 & 4.10 & 0.420 & 0.193 & 0.0776\\
OH94 thin mantle, 0~yr & 19.3 (14.0) & 11.8 (9.57) & 8.56 (7.77) & 6.89 (6.77) & 4.88 & 0.762 & 0.389 & 0.180\\
OH94 thin mantle, $10^5$yr, $10^6\:{\rm{cm^{-3}}}$ & 24.7 (17.9) & 14.6 (11.9) & 10.4 (9.43) & 8.24 (8.09) & 6.86 & 1.14 & 0.603 & 0.290\\
OH94 thin mantle, $10^7$yr, $10^6\:{\rm{cm^{-3}}}$ & 26.5 (19.2) & 16.2 (13.2) & 11.7 (10.6) & 9.18 (9.01) & 8.33 & 1.42 & 0.746 & 0.356\\
OH94 thick mantle, 0~yr & 36.0 (26.1) & 15.1 (12.3) & 16.7 (15.1) & 9.34 (9.17) & 10.6 & 3.13 & 0.950 & 0.290\\
OH94 thick mantle, $10^5$yr, $10^6\:{\rm{cm^{-3}}}$ & 43.5 (31.5) & 17.8 (14.5) & 18.7 (17.0) & 10.7 (10.5) & 13.2 & 3.96 & 1.27 & 0.404\\
OH94 thick mantle, $10^7$yr, $10^6\:{\rm{cm^{-3}}}$ & 45.7 (33.1) & 18.2 (14.8) & 19.4 (17.6) & 10.9 (10.7) & 14.8 & 4.55 & 1.44 & 0.450\\
\enddata
\tablenotetext{a}{Common total to dust mass ratio of 142 is adopted (Draine 2011).}
\tablenotetext{b}{References: WD01 - \citet{weingartner2001}; OH94 - \citet{ossenkopf1994}, opacities have been scaled from values in parentheses to include contribution from scattering.}
\tablenotetext{c}{Mean wavelengths weighted by filter response and background spectrum.}
\label{tb:kappa}
\end{deluxetable}

\clearpage
\section{Results}

\subsection{Saturated regions and measurement of the spectrum of the Galactic foreground}


After a global investigation of the IRDC, we focus on three
1$\arcmin\times$1$\arcmin$ regions around BT12 and BTK core/clumps, C1
(Fig.~\ref{fig:c1}), C4 (Fig.~\ref{fig:c4}) and C11
(Fig.~\ref{fig:c11}) to search for local saturation at 8, 24, and
70\micron.
As seen in the top panels of these figures, the 8\micron\ saturation
is more widespread than seen by BT12, i.e. in C1 and C4, which is a
consequence of searching in local regions and thus reducing the
masking effect of foreground variations. There is generally good
correspondence of saturated regions across the different wavebands,
although this can be disrupted by the presence of discrete
sources. There is a tendency for saturation to be more extended at 8
and 24\micron, than at 70\micron\ (however, not in C11). The size of
the $2\sigma$ saturated region in C1 at 70\micron\ is not greater
than 12\arcsec\, but it is larger if the condition is relaxed to
$4\sigma$, so we consider this likely to be a saturated core. In
general, comparing the 2, 4, 8$\sigma$-defined saturated regions, we
see coherent, contiguous morphologies, which indicates that these
levels are revealing real cloud structures and that the image noise
levels are reasonably well-estimated.

We derive the specific intensities of the Galactic foreground 
towards C1, C4, C11, also including measurements in IRAC bands 1 to 3
at the locations of 8\micron\ pixels showing $2\sigma$-saturation. We
plot these intensities, normalized by the 8\micron\ values, on
Figure~\ref{fig:wholecloud}b.
In general, these measurements agree well with the LD01 model. The
24\micron\ values are consistently about a factor of two smaller,
while the 70\micron\ values are a few tens of percent larger. These
particular ratios are sensitive to our choice of normalizing at
8\micron. In the IRAC bands 1 and 2, the observed intensities are
significantly lower than the LD01 model of total emission. However,
there is reasonable agreement with just the dust component of Galactic
emission (dotted line). This IRDC is at 5~kpc distance, and this
appears to be close enough that foreground stars are not making
significant contributions to the foreground emission as measured on
$\sim$arcsecond scales in these saturarted cores.


\subsection{FIR Extinction Maps}

Following the method described in \S\ref{S:methods}, given estimates
of the foreground and background intensities towards each region we
derive the extinction maps at 8, 24 and 70\micron, displaying $\Sigma$
in $\rm{g\:cm^{-2}}$ under the assumption of opacities of the thick ice
mantle model of OH94 (middle rows of Figs.~\ref{fig:c1}, \ref{fig:c4},
\ref{fig:c11}). The choice of this thick, rather than thin, ice mantle
dust model is motivated by the observed opacity law (below), and is
different from BT12's use of the thin ice mantle model. For
8\micron-derived maps the resulting variation of opacity per unit mass
and thus $\Sigma$ is quite small, $\sim30\%$, but it can make more
than a factor of 3 difference at 70\micron.

The value of $\Sigma$ at which our saturation-based extinction mapping
begins to start underestimating the true mass surface density is (BT12)
\begin{equation}
\Sigma_{\rm{sat}}=\frac{\tau_{\nu,{\rm{sat}}}}{\kappa_{\nu}}=\frac{{\rm{ln}}(I_{\nu,0}/I_{\nu,1})}{\kappa_{\nu}},
\end{equation}
where $I_{\nu,1}$ is set equal to the $2\sigma$ noise level, i.e. 1.2,
0.5, 30~MJy/sr for the 8, 24, 70\micron\ images. At these wavelengths,
the saturated cores have average values of $I_{\nu,0}=$92.7, 53.7,
1302.3~MJy/sr, respectively. Thus for thick ice mantle opacities,
$\Sigma_{\rm{sat}}=0.44,0.39,1.05\:{\rm{g\:cm^{-2}}}$, respectively,
while for thin ice mantle opacities it is
$0.58,0.75,3.66\:{\rm{g\:cm^{-2}}}$. Note, as discussed by BT12,
higher values of $\Sigma$ than $\Sigma_{\rm{sat}}$ will be present in
the map, but these will be tend to be more and more affected by
saturation, leading to underestimation of the true $\Sigma$ values.

The above analysis predicts that we should be able to probe to higher
values of $\Sigma$ in the 70\micron\ maps than in the 8 and
24\micron\ maps, and that such high $\Sigma$ structures will be in
regions that have been identified as being saturated at the two
shorter wavelengths. In core/clump C1, we indeed see this morphology:
the 70\micron\ map peaks at around $\rm{1\:g\:cm^{-2}}$ at a position
close to the C1 core center identified by BT12 and BTK. This highest
column density region is coincident with the C1-N $\rm{N_2D^+}(3-2)$
core identified by Tan et al. (2013b). We also infer that the
70\micron\ map is revealing real structures in the range
$\Sigma\sim0.5-1\:{\rm{g\:cm}^{-2}}$, which are not so reliably probed
by the shorter wavelength maps.

In the C4 region we see two main high $\Sigma$ peaks that exhibit
local saturation, with the western structure identified as C4 by BT12
and the eastern as C13 by BTK. In between are two sources seen most
clearly at 24\micron. These sources undoubtedly affect the MIREX and
FIREX methods in this region, so it is possible that there is a high
$\Sigma$ bridge between the two cores.

In C11, there is a relatively widespread high $\Sigma$ region, which
is part of a highly filamentary structure extending towards C12 to the
NE (BTK). The 70\micron\ map also reveals a relatively high $\Sigma$
region extending to the NW. This extension is seen in the 8 and
24\micron\ maps, but at lower $\Sigma$. It is possible that localized
foreground variations at the shorter wavelengths are affecting the 8
and 24\micron\ maps (see also BTK, where this extension is more
prominent in an 8\micron\ map derived with a finer decomposition of
foreground variations).


\subsection{The MIR to FIR Extinction Law and Evidence for Grain Growth}

\begin{figure}
\begin{center}$
\begin{array}{ccc}
\hspace{-0.1in} \includegraphics[width=2.15in]{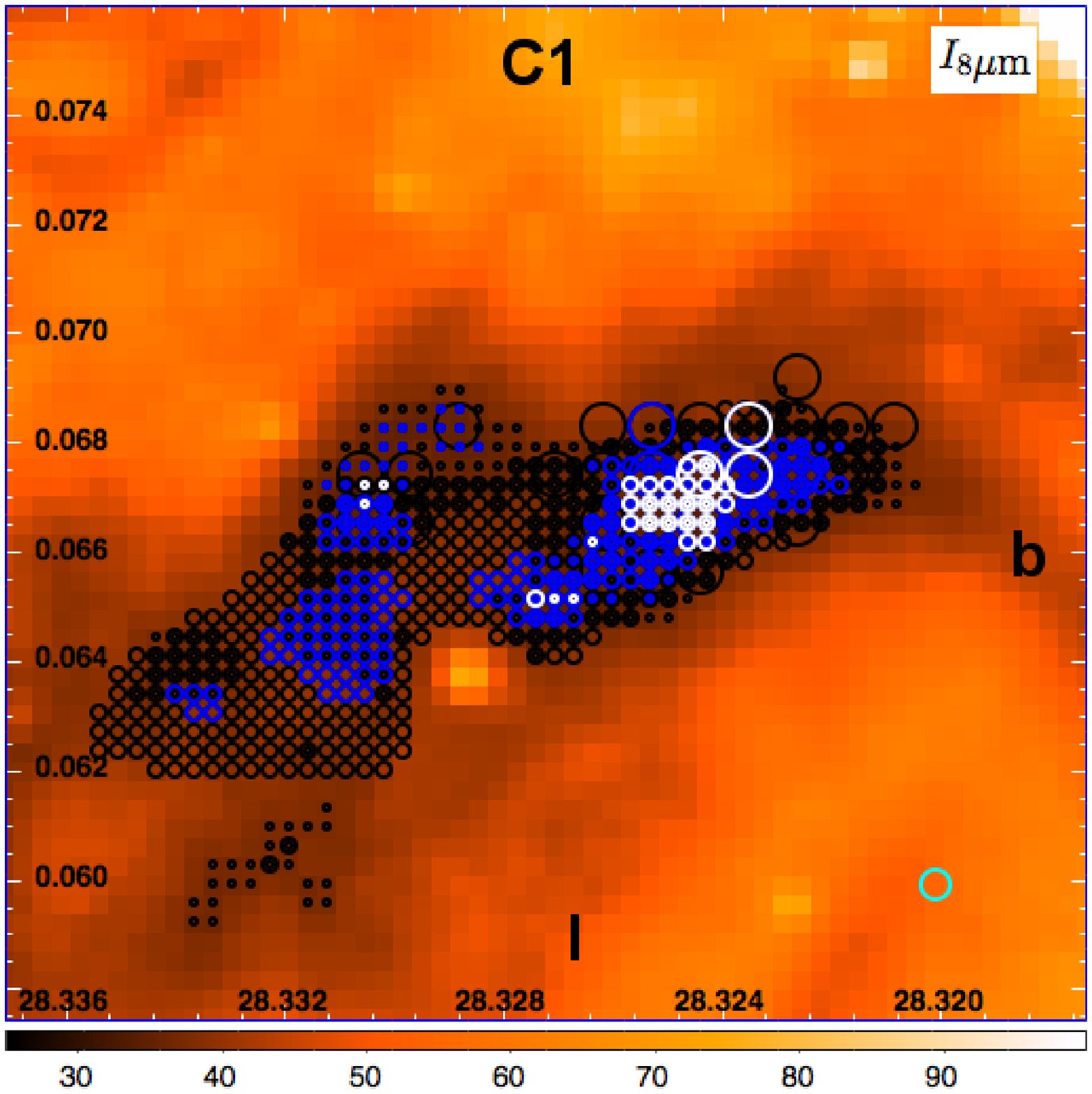} & \hspace{-0.1in} \includegraphics[width=2.15in]{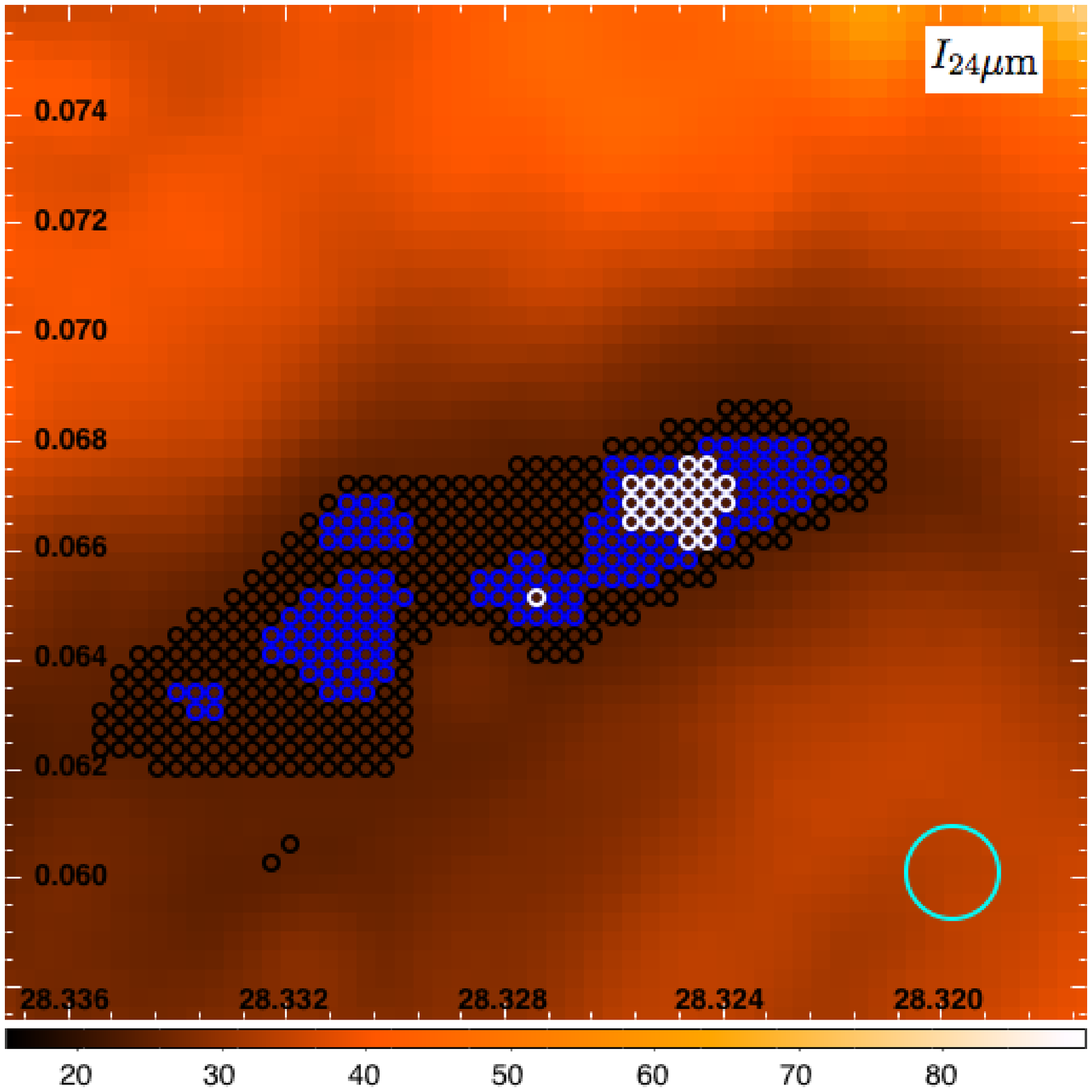} & \hspace{-0.1in} \includegraphics[width=2.15in]{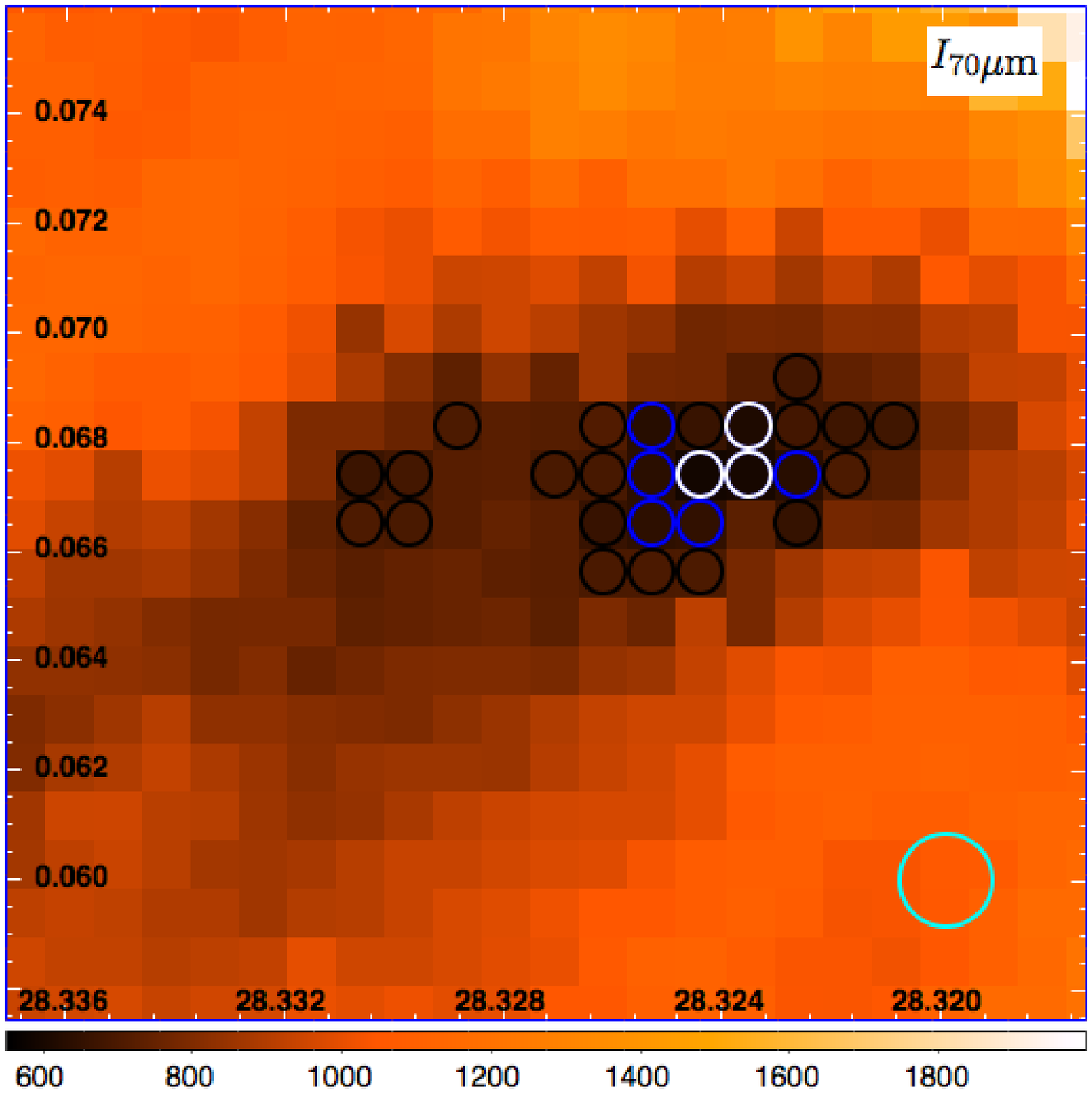} \\
\hspace{-0.1in} \includegraphics[width=2.15in]{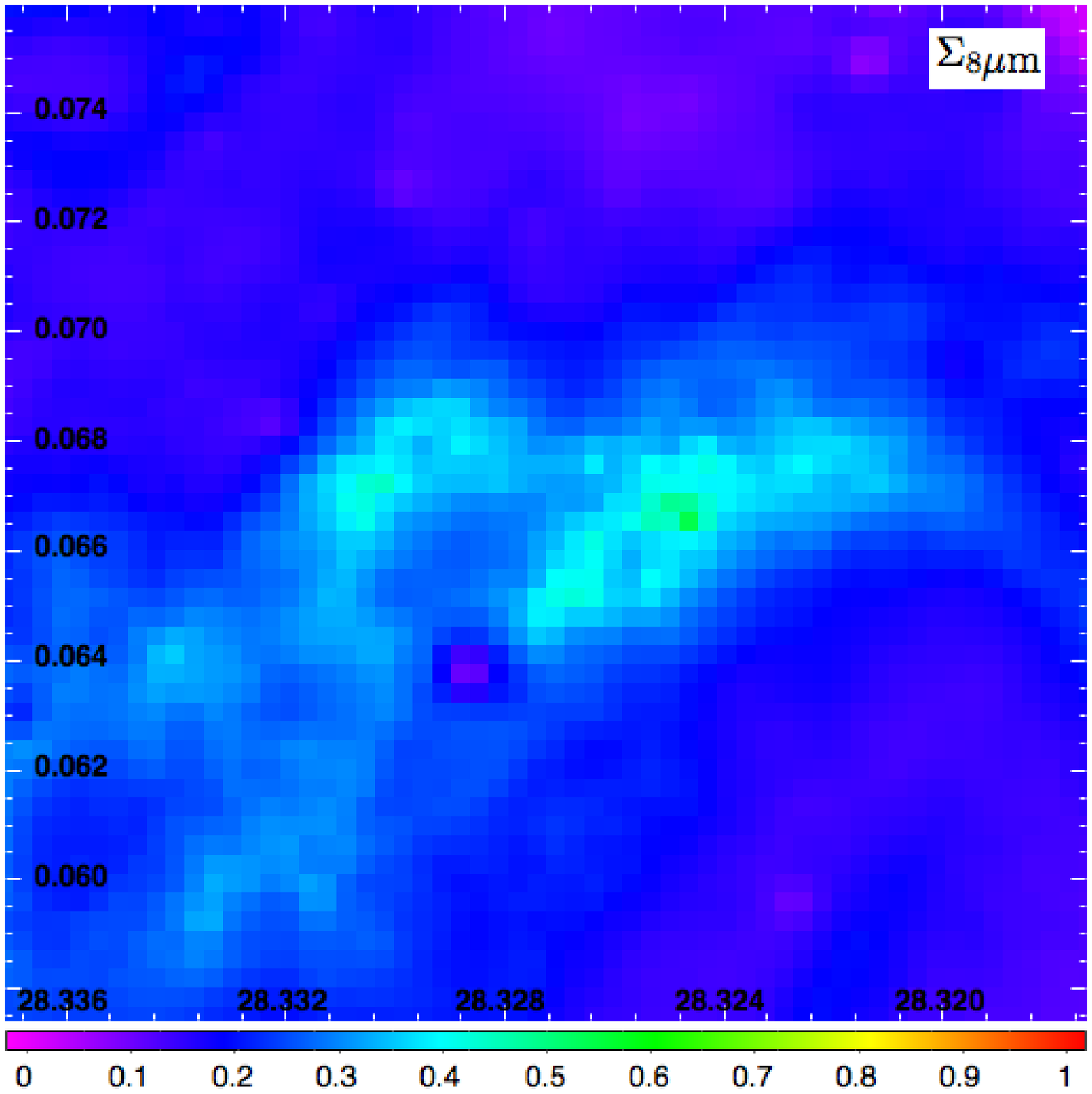} &\hspace{-0.1in} \includegraphics[width=2.15in]{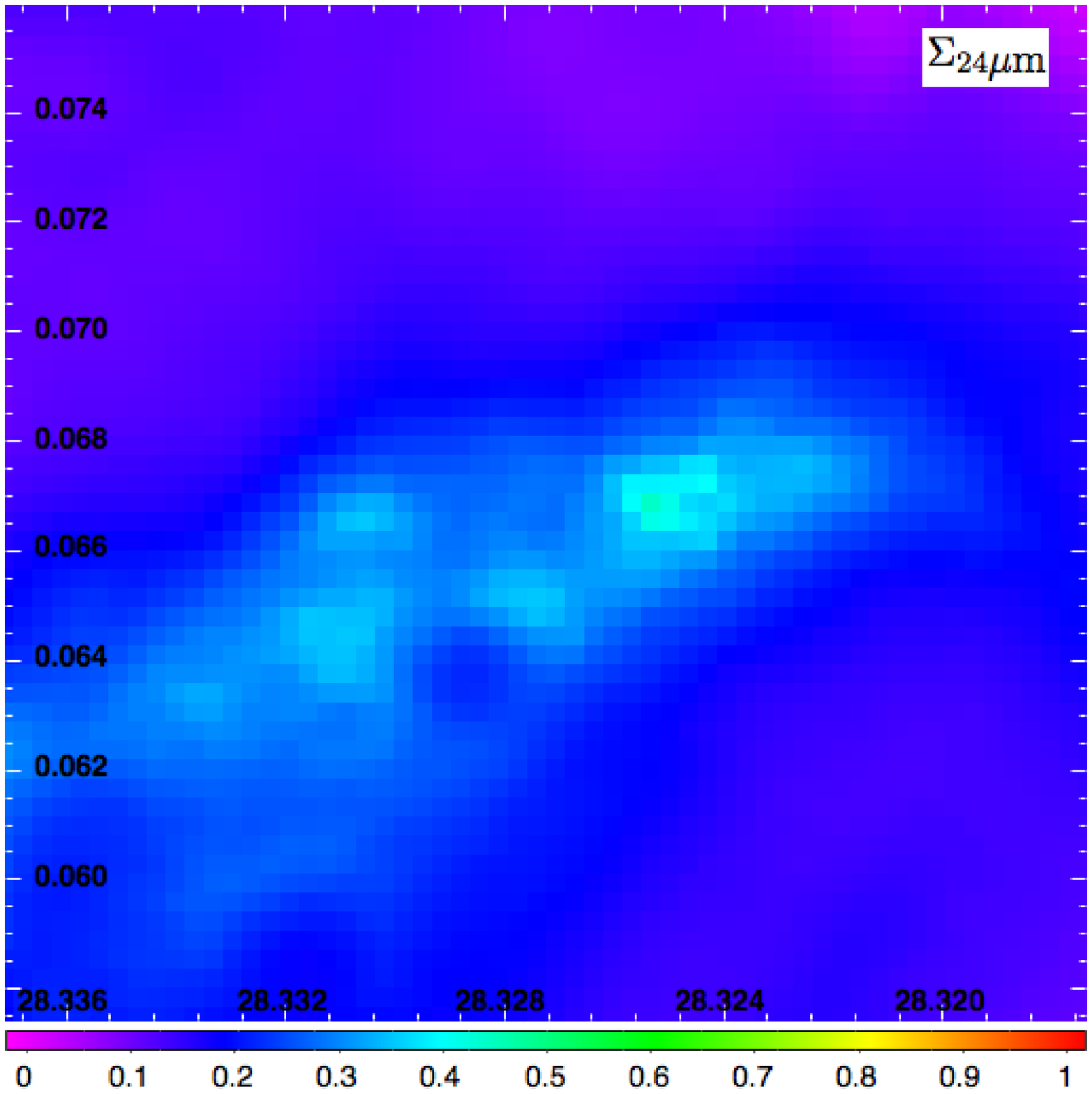} &\hspace{-0.1in} \includegraphics[width=2.15in]{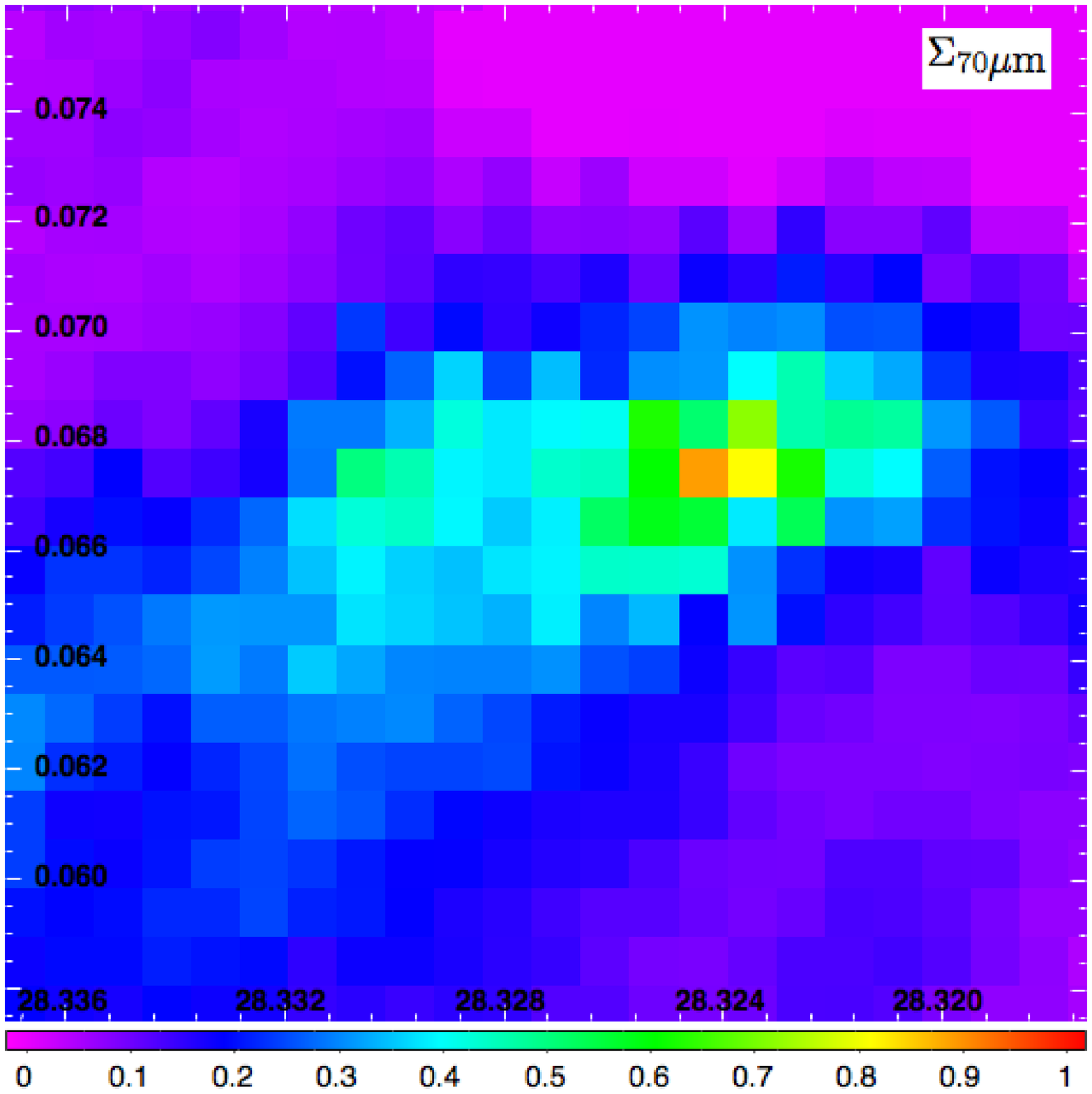} \\
\hspace{-0.1in} \includegraphics[width=2.15in]{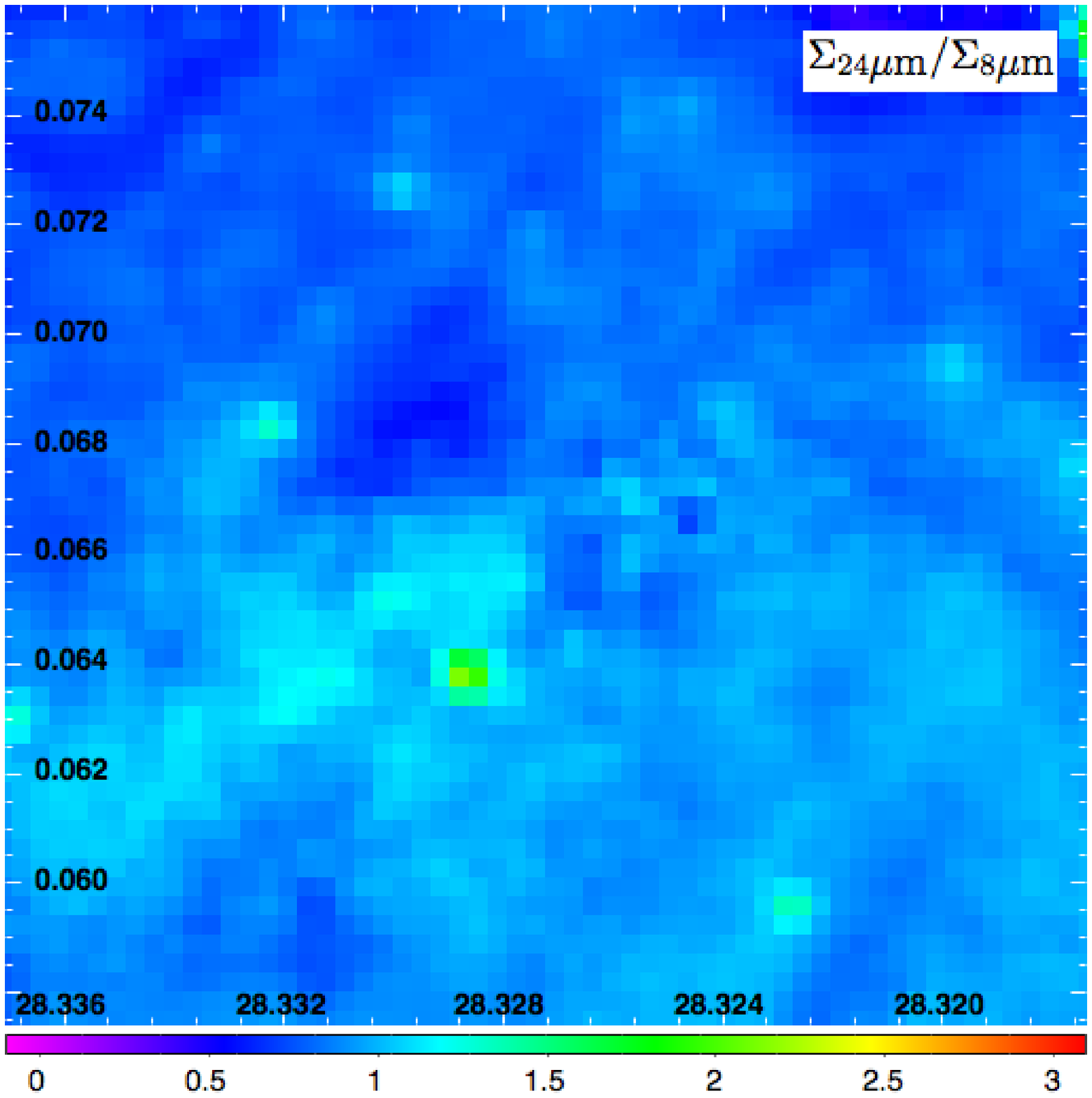} &\hspace{-0.1in} \includegraphics[width=2.15in]{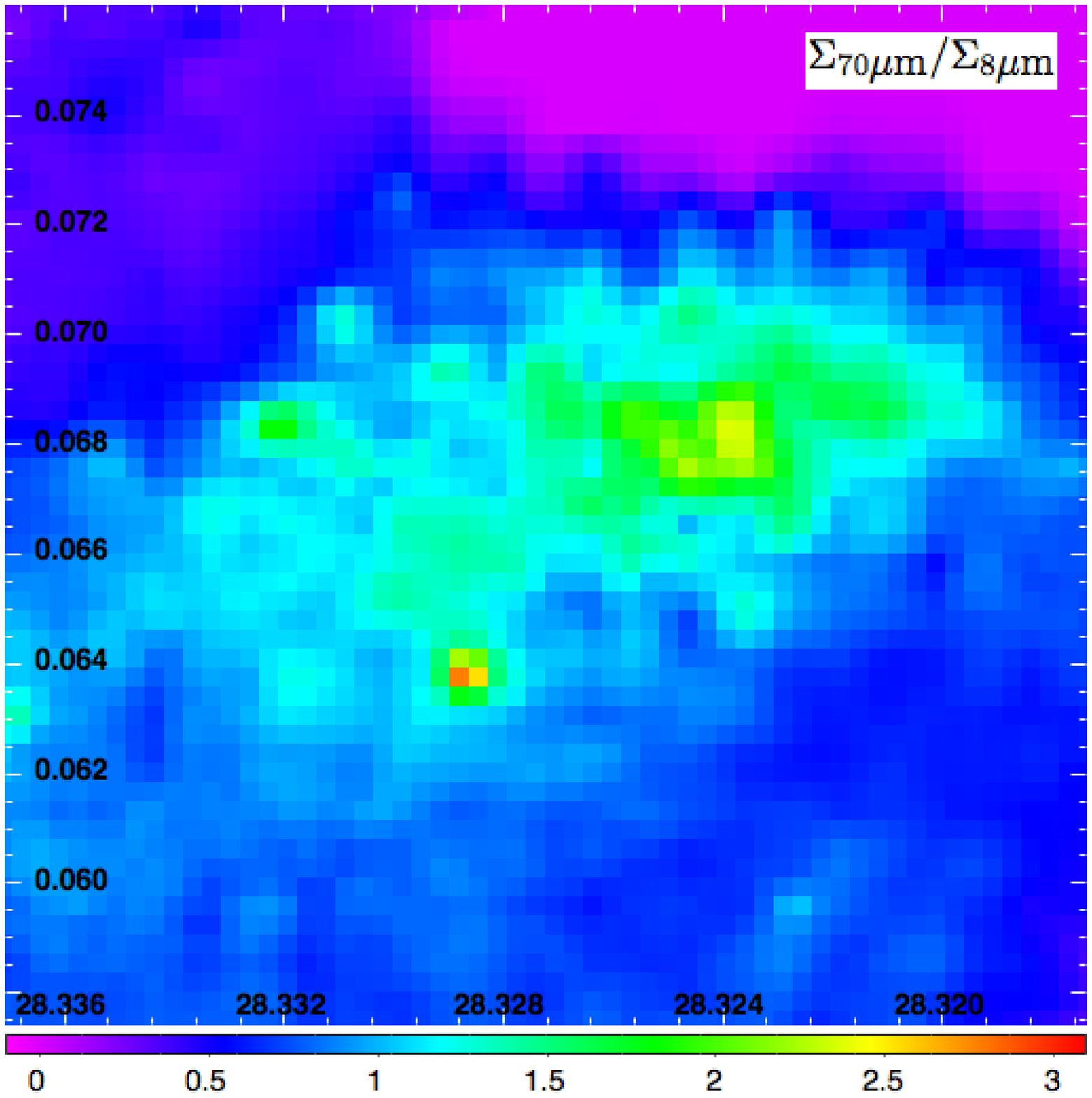} &\hspace{-0.1in} \includegraphics[width=2.15in]{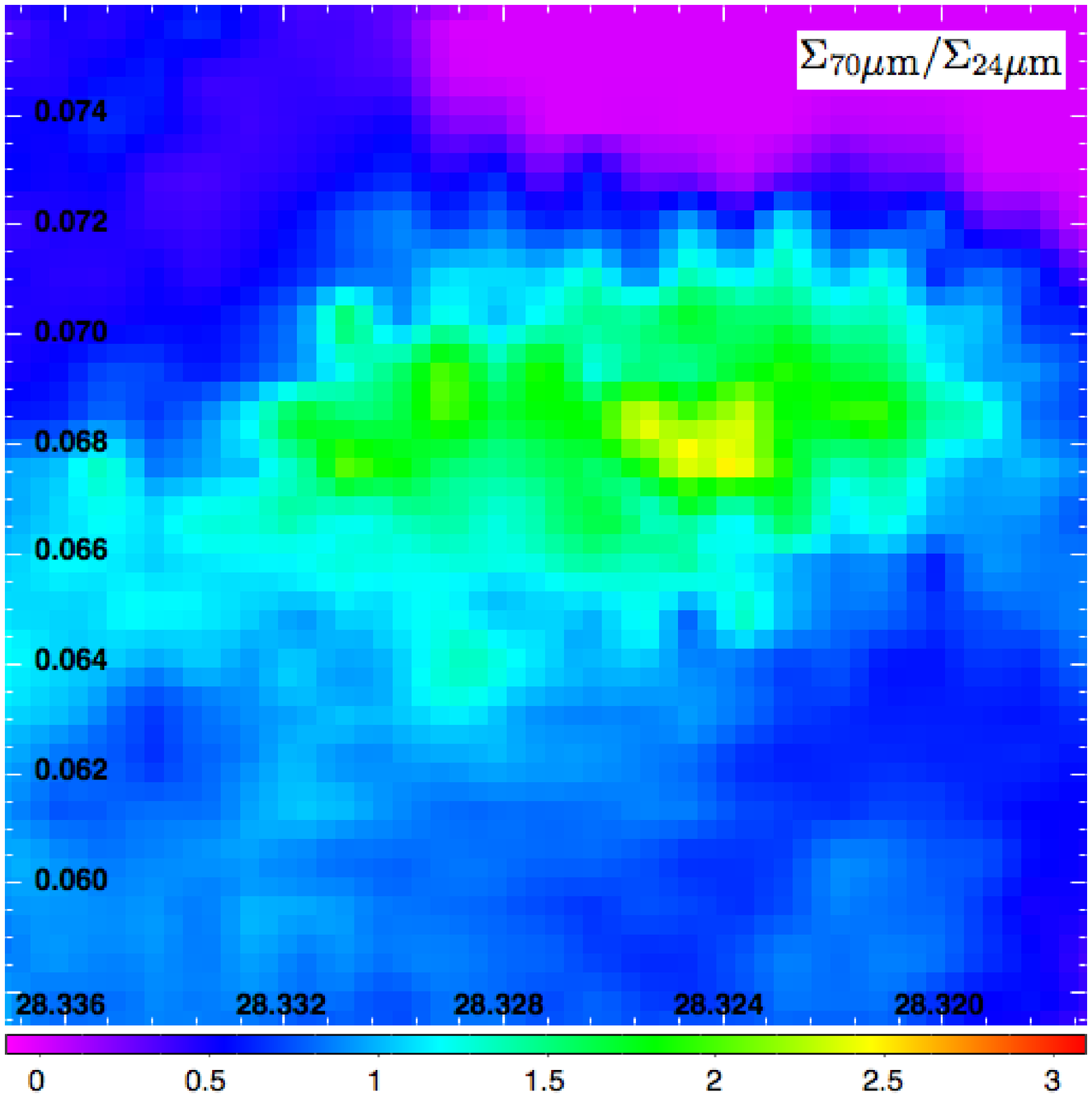} \\
\end{array}$
\end{center}
\caption{
\footnotesize {\it Top row:} Saturated regions in IRDC C1 viewed in
Galactic coordinates at 8, 24 \& 70\micron\ (left, middle, right
panels; intensity scales in MJy/sr). We have searched inside the
displayed 1$\arcmin\times1\arcmin$ square for saturated intensities
in each pixel, defined to be within 2/4/8$\sigma$ (white/blue/black,
respectively) of the minimum pixel value. Small/medium/large circles
show 8/24/70\micron\ saturation, respectively, which are displayed
together on the left panel and separately for 24 and 70\micron\ on the
middle and right panels.
The circles in the lower-right indicate the beam. 
{\it Middle row:} $\Sigma$ maps derived from the 8, 24,
70\micron\ images (left, middle, right panels; intensity scale in $\rm
g\:cm^{-2}$) assuming the moderately coagulated thick ice mantle dust
model of OH94.
{\it Bottom row:} Ratios of $\Sigma$ maps (or equivalently deviation of
relative opacity from the OH94 thick ice mantle dust model) for
$\Sigma_{24\micron}/\Sigma_{8\micron}$ (left),
$\Sigma_{70\micron}/\Sigma_{24\micron}$ (middle) and
$\Sigma_{70\micron}/\Sigma_{8\micron}$ (right). 
}
\label{fig:c1}
\end{figure}

\begin{figure}
\begin{center}$
\begin{array}{ccc}
\hspace{-0.1in} \includegraphics[width=2.15in]{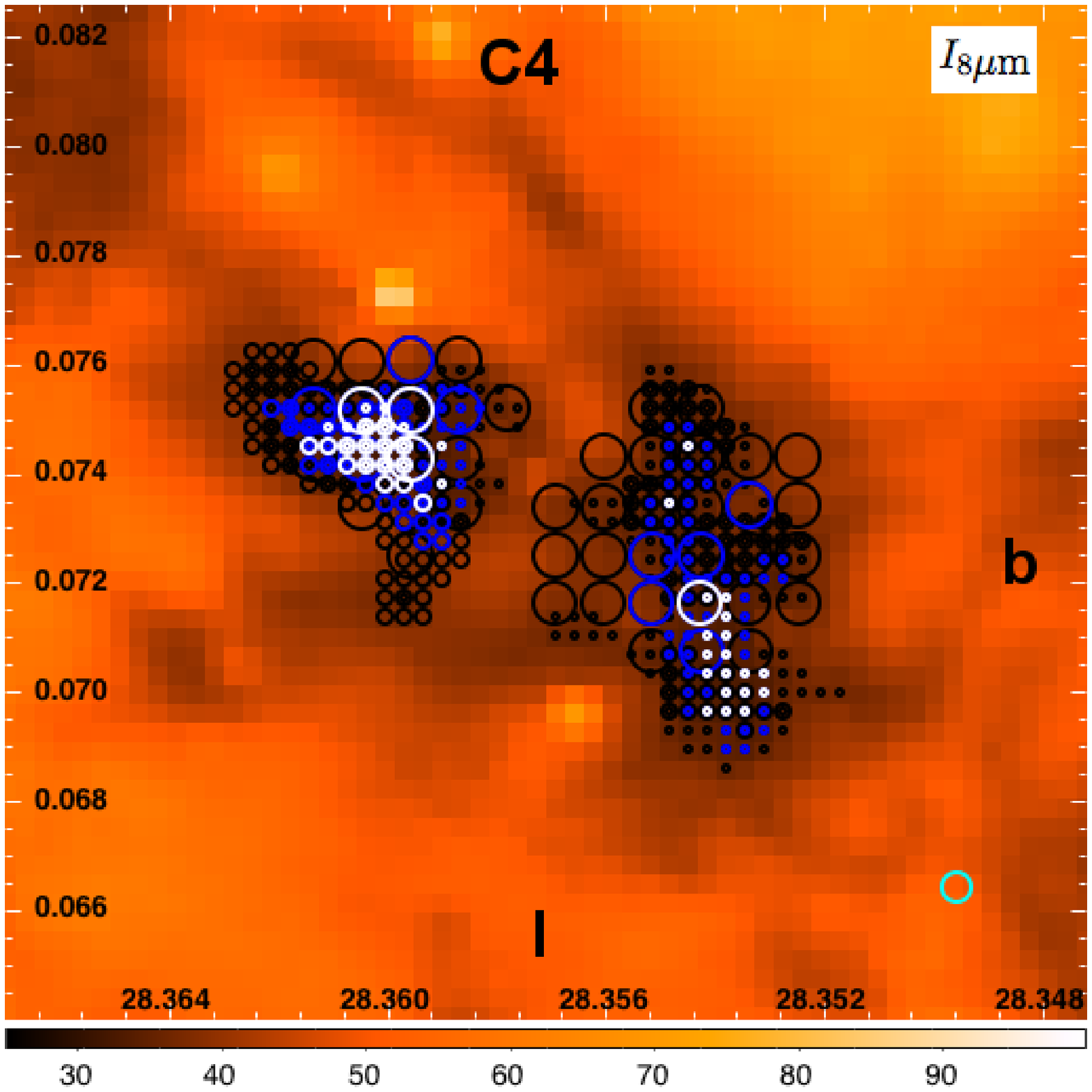} & \hspace{-0.1in} \includegraphics[width=2.15in]{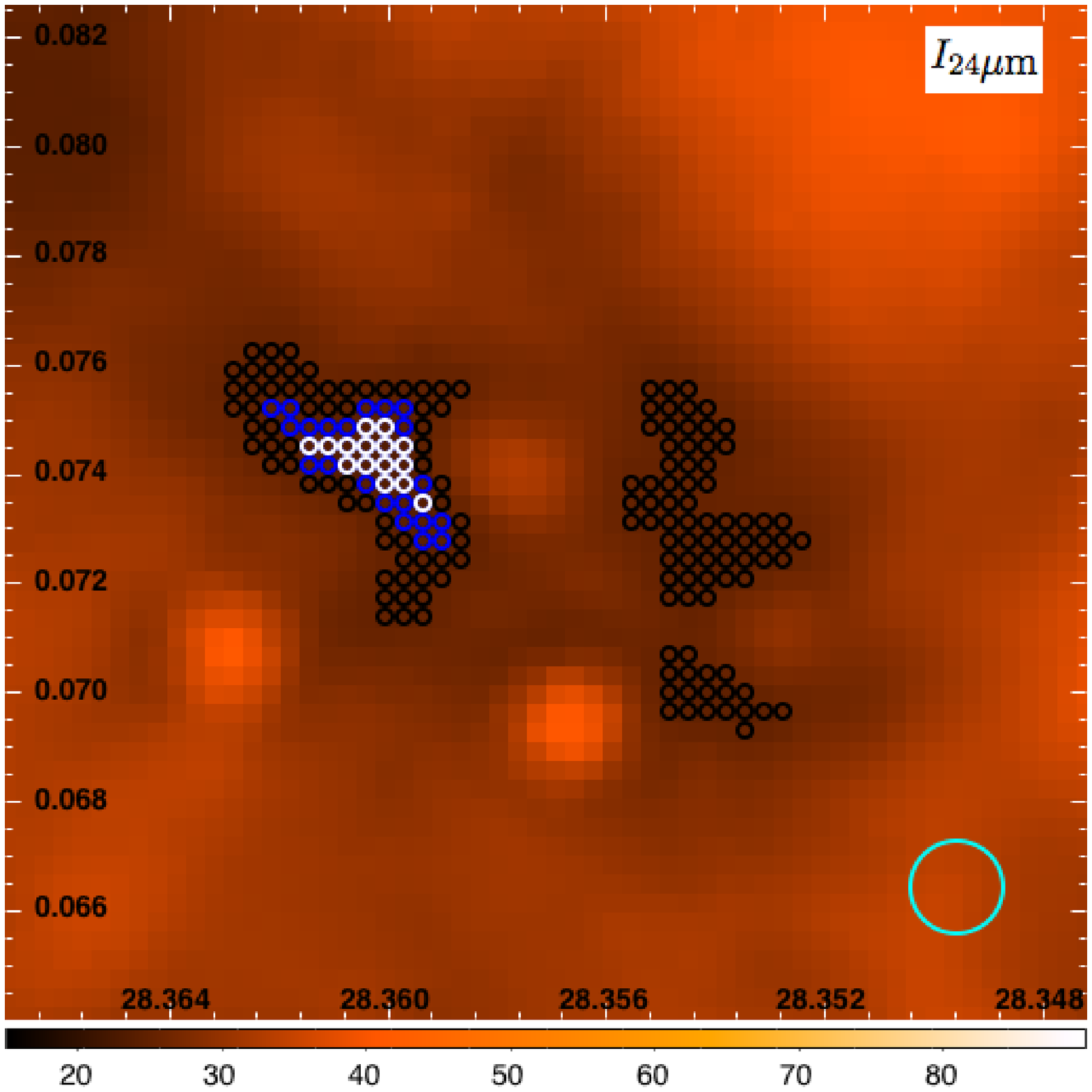} & \hspace{-0.1in} \includegraphics[width=2.15in]{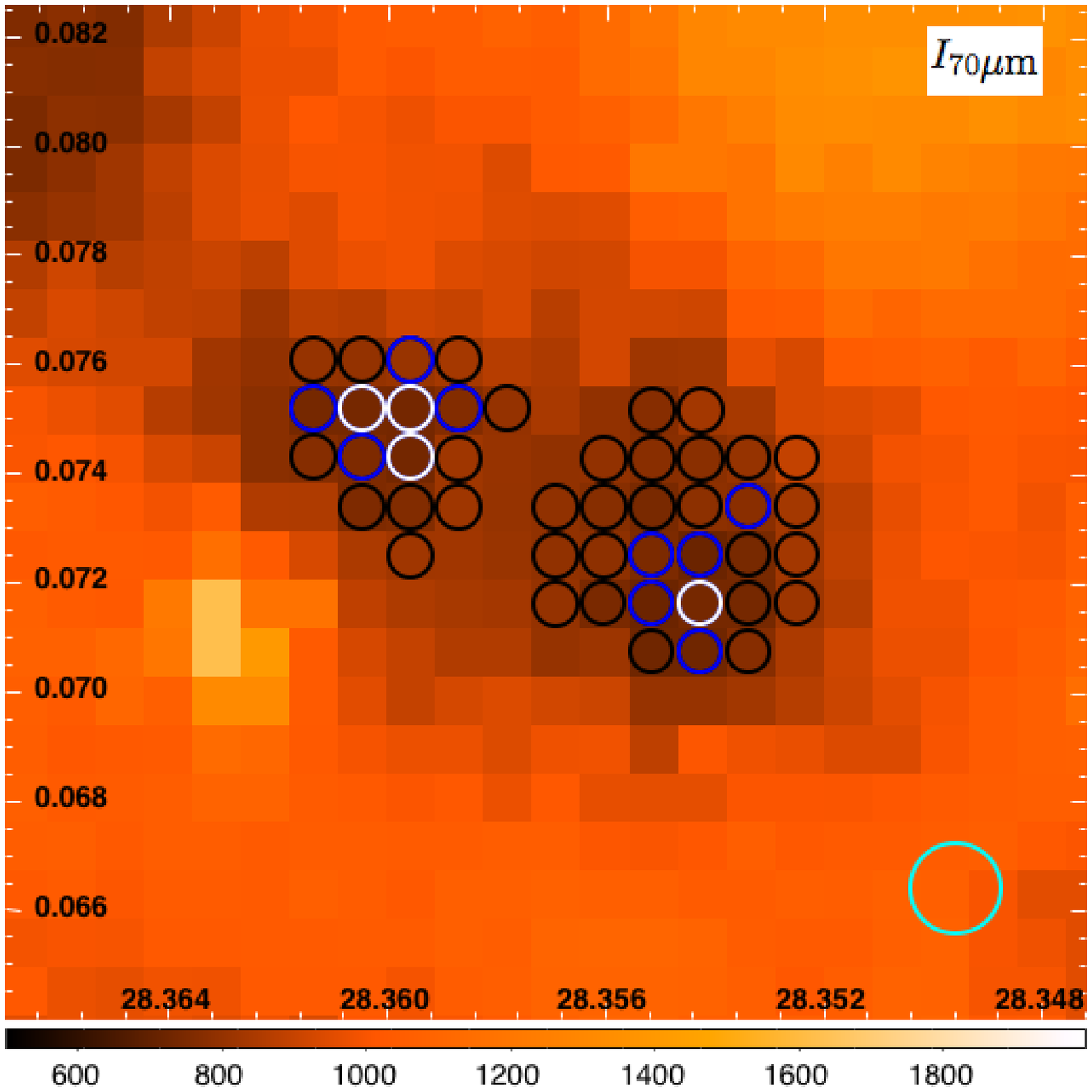} \\
\hspace{-0.1in} \includegraphics[width=2.15in]{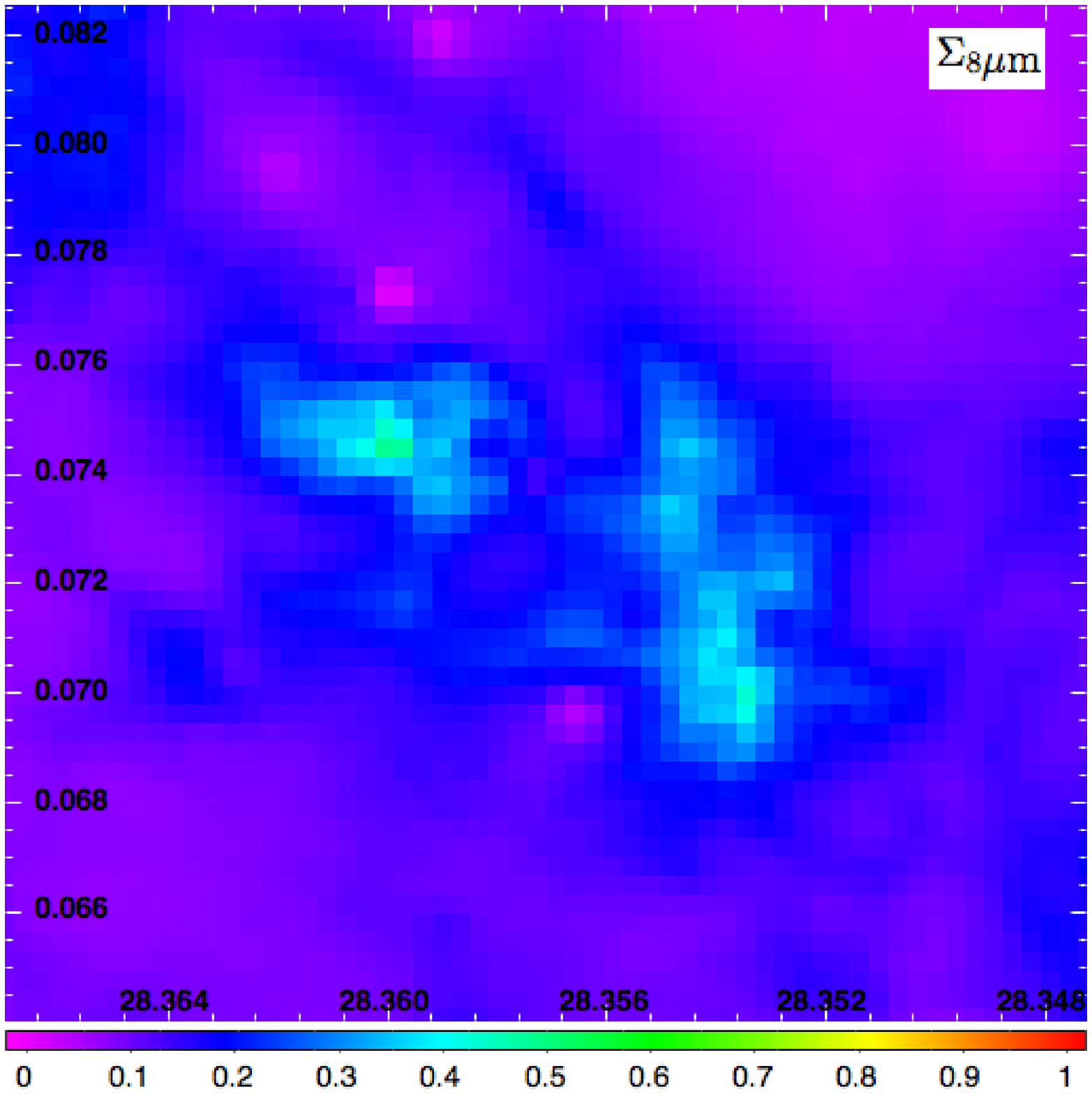} &\hspace{-0.1in} \includegraphics[width=2.15in]{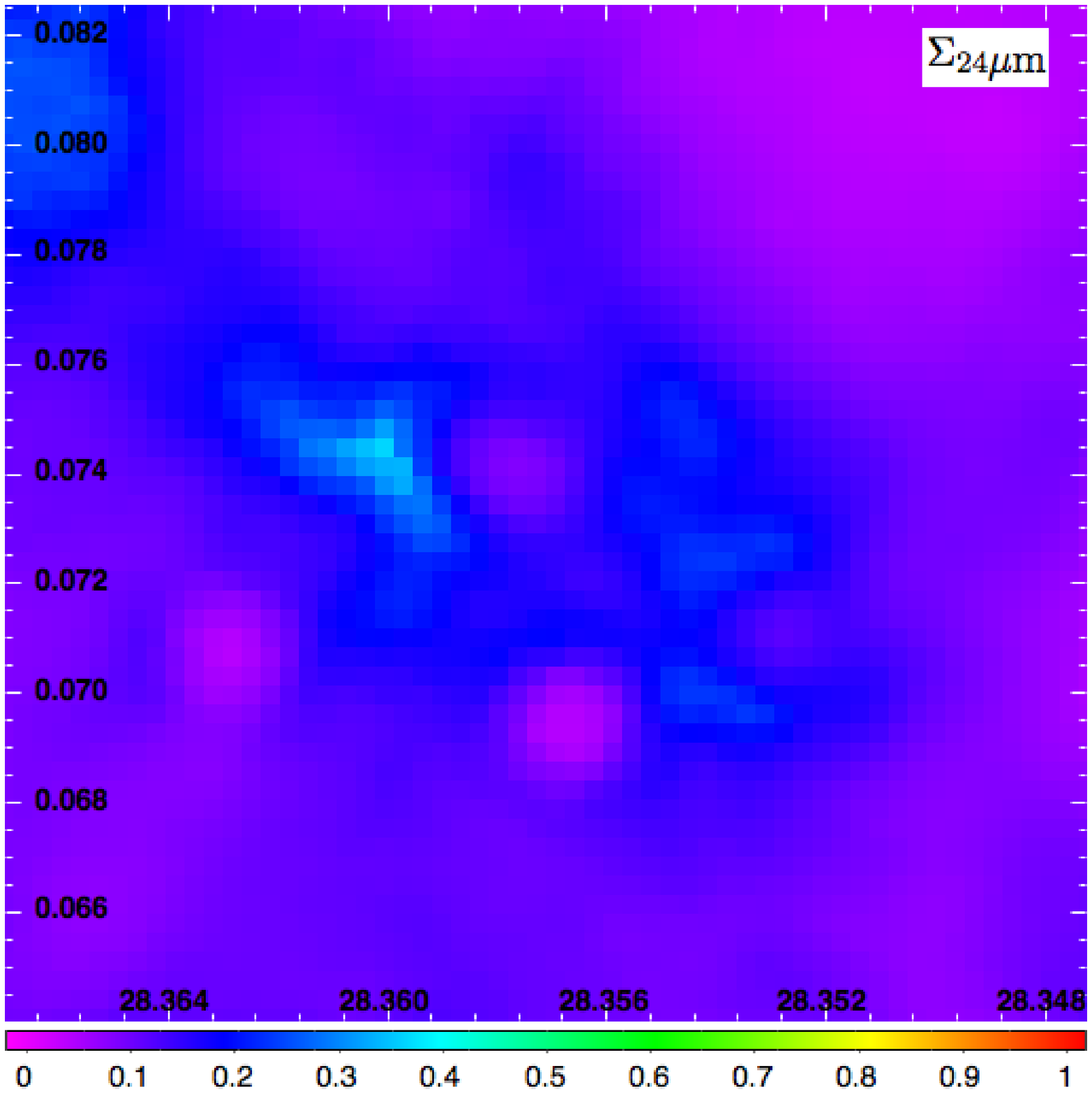} &\hspace{-0.1in} \includegraphics[width=2.15in]{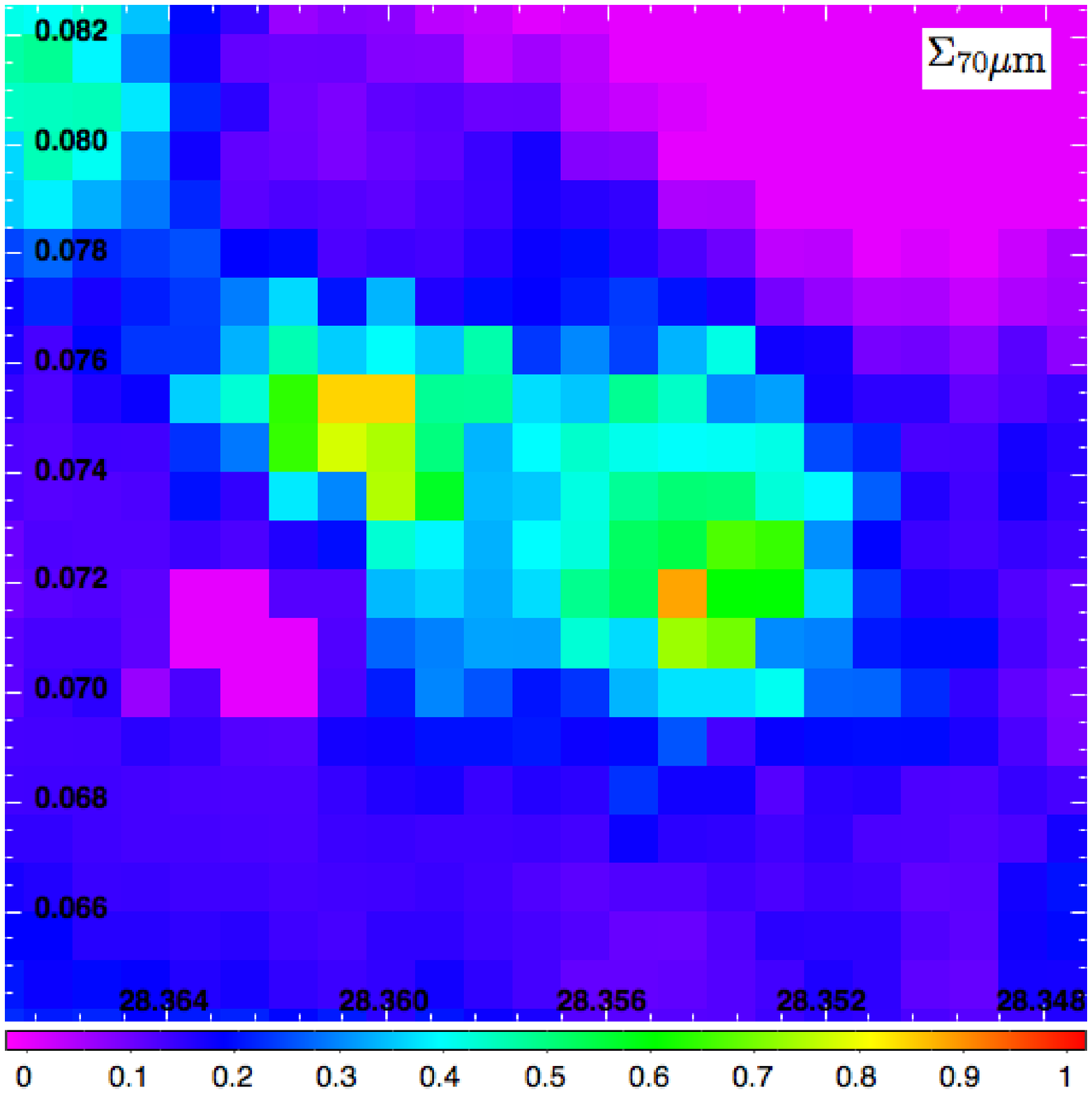} \\
\hspace{-0.1in} \includegraphics[width=2.15in]{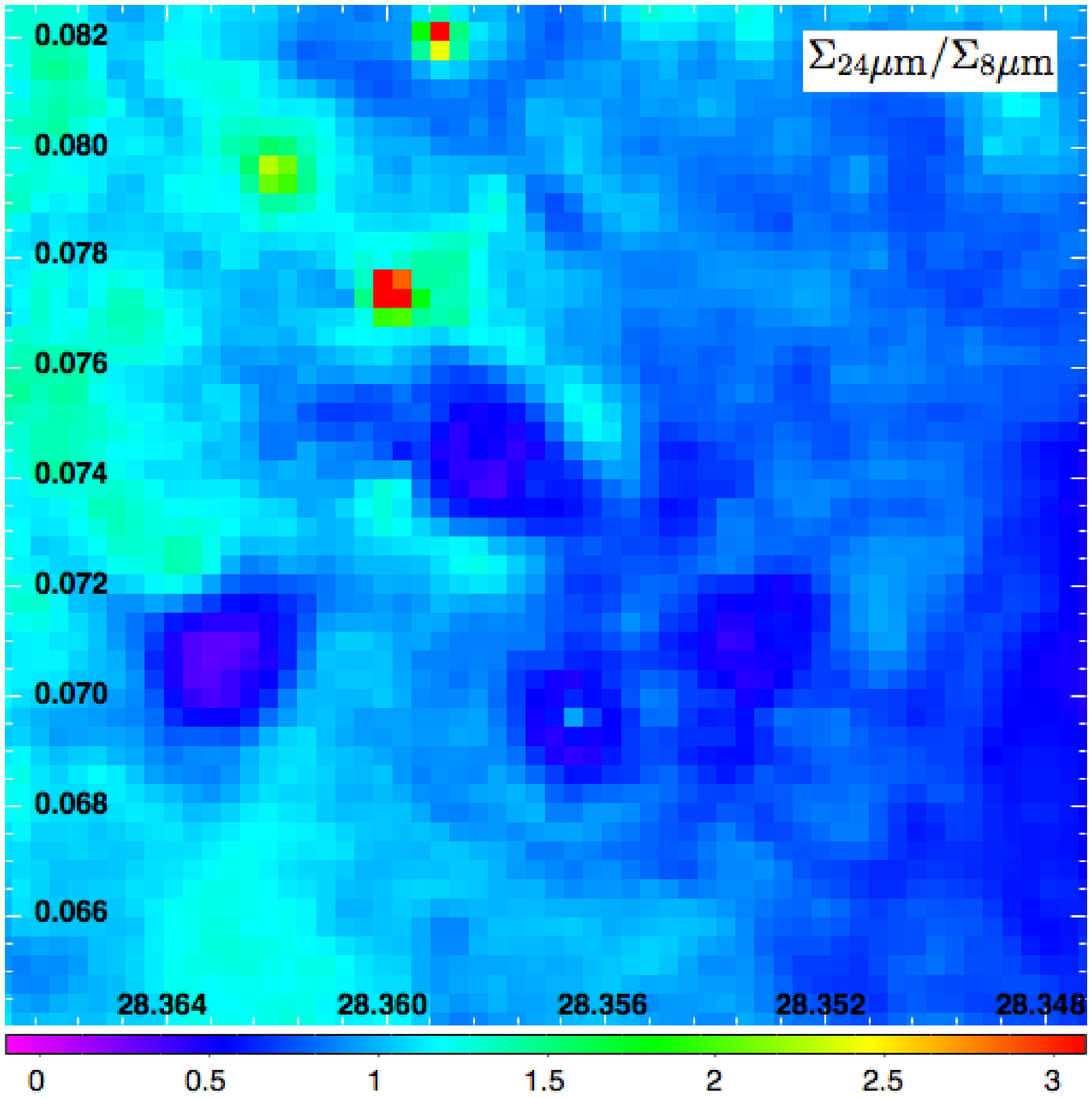} &\hspace{-0.1in} \includegraphics[width=2.15in]{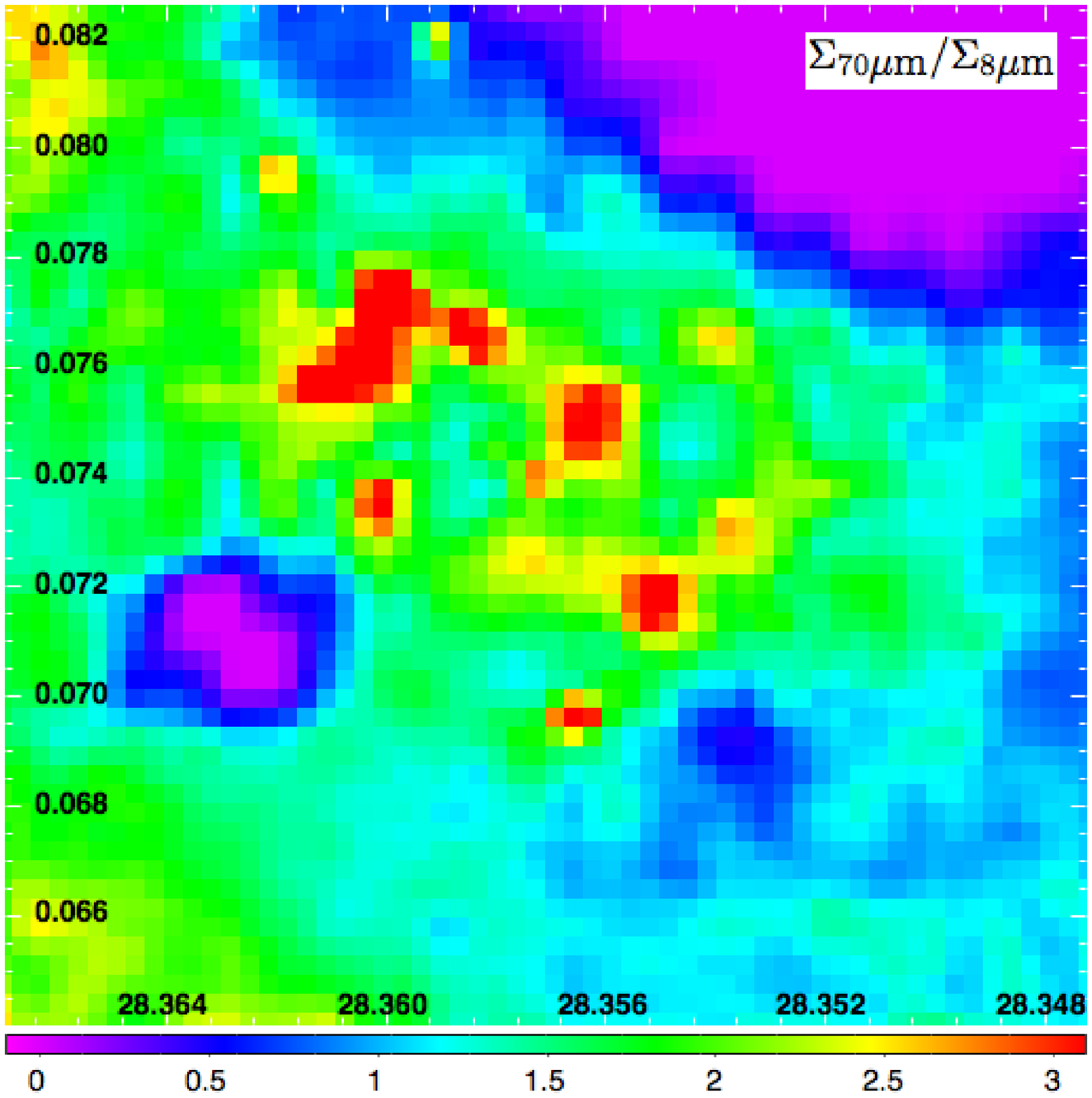} &\hspace{-0.1in} \includegraphics[width=2.15in]{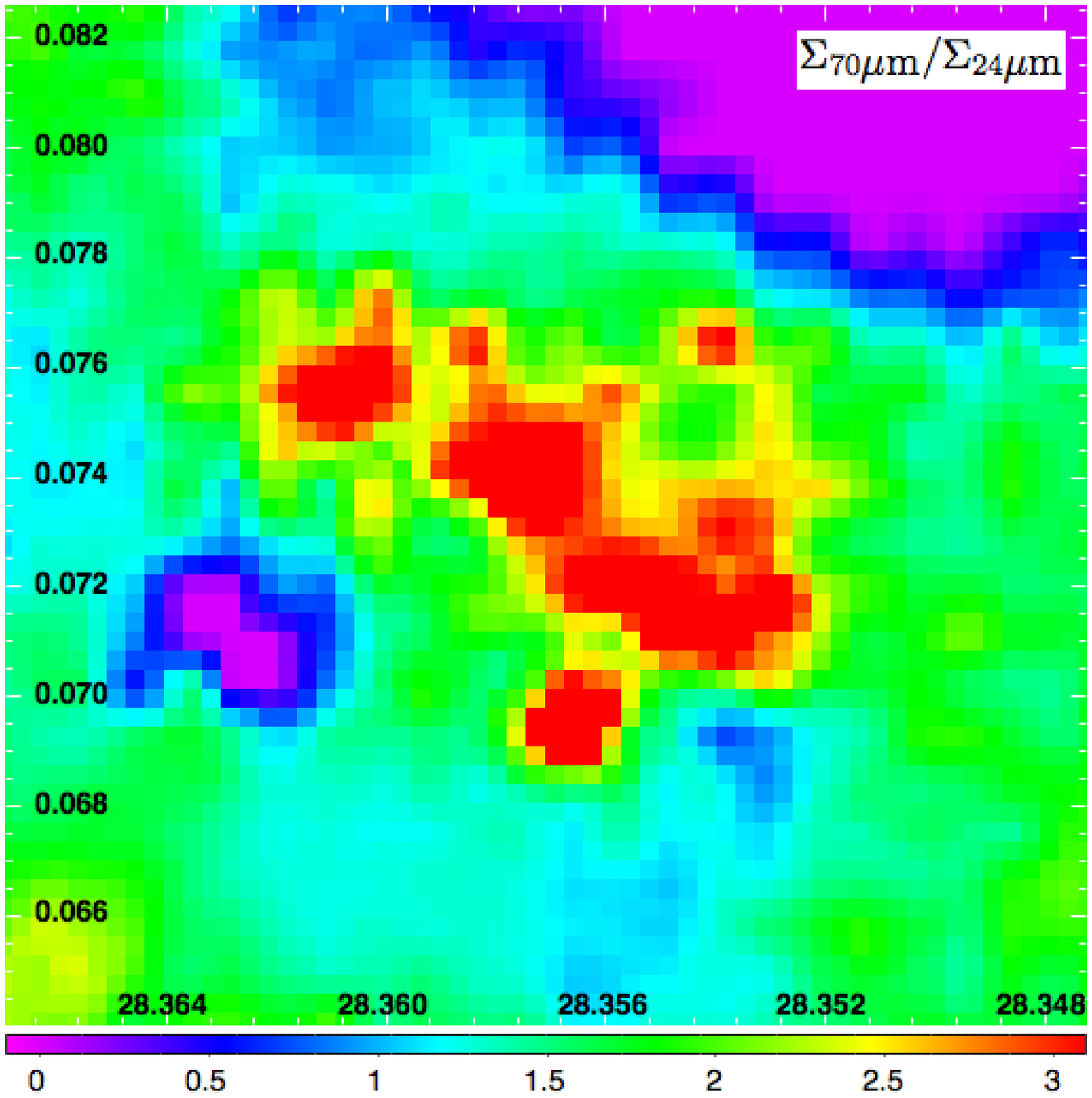} \\
\end{array}$
\end{center}
\caption{\footnotesize Same with Figure~\ref{fig:c1} but for C4.}
\label{fig:c4}
\end{figure}

\begin{figure}
\begin{center}$
\begin{array}{ccc}
\hspace{-0.1in} \includegraphics[width=2.15in]{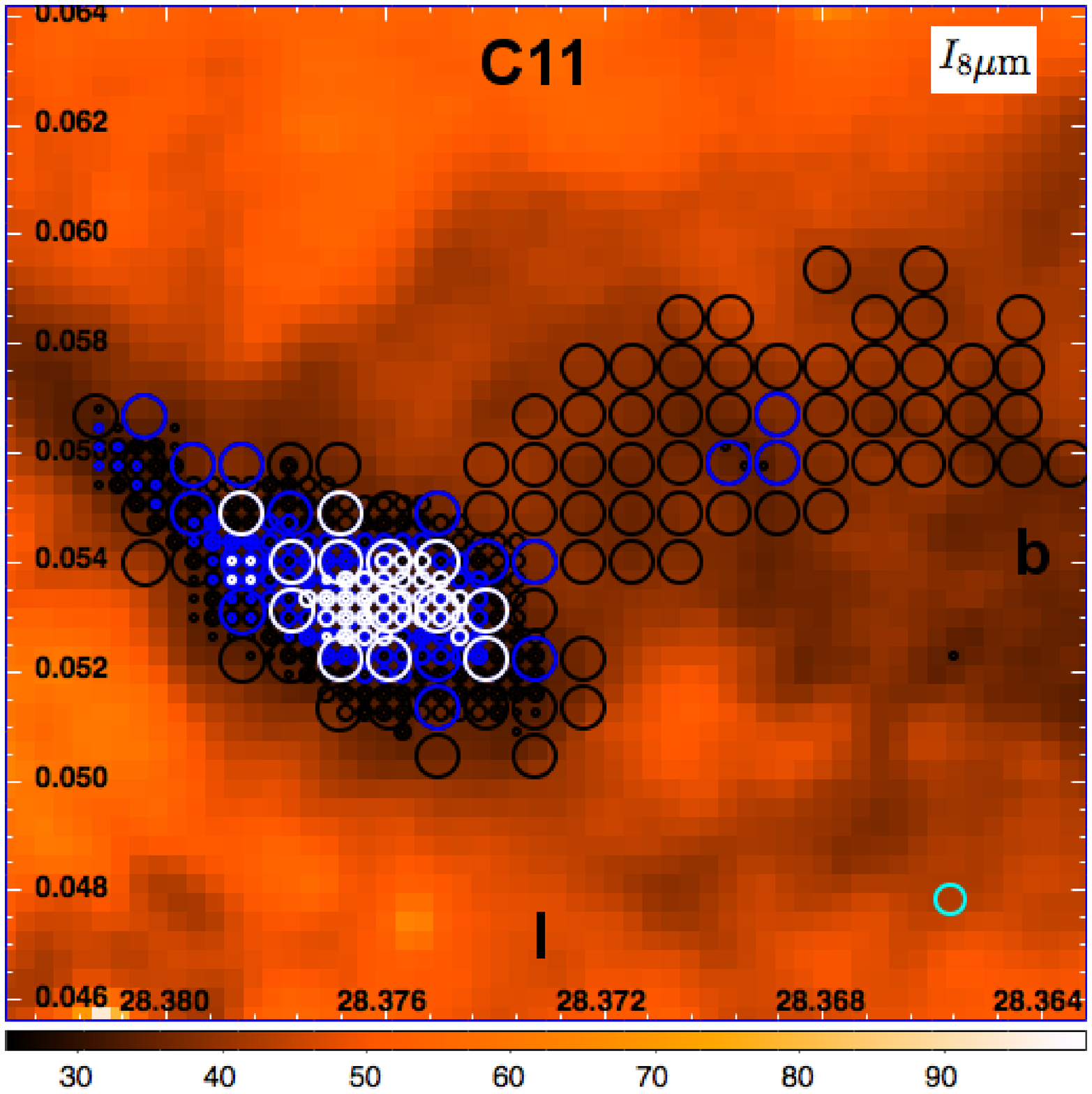} & \hspace{-0.1in} \includegraphics[width=2.15in]{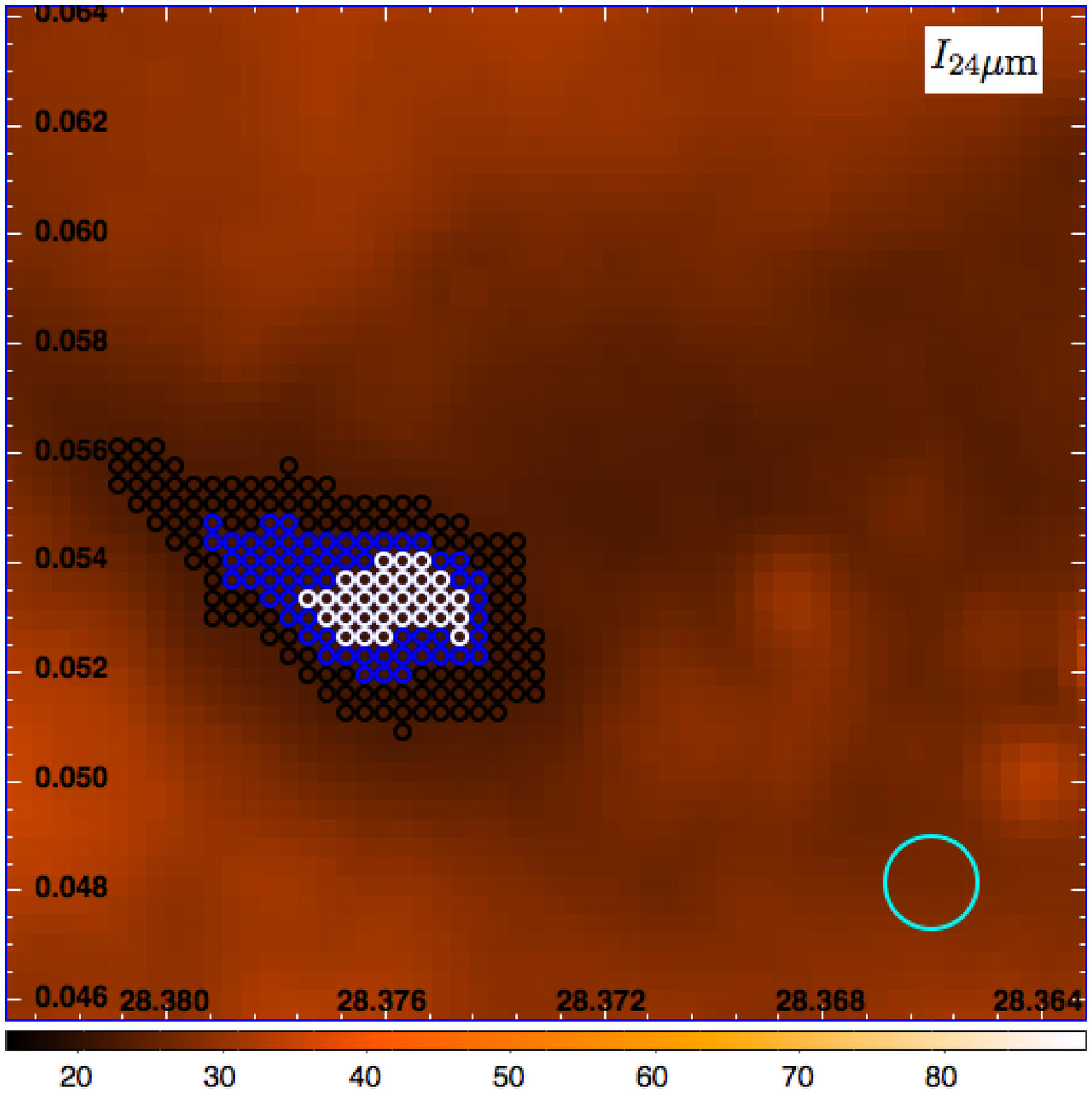} & \hspace{-0.1in} \includegraphics[width=2.15in]{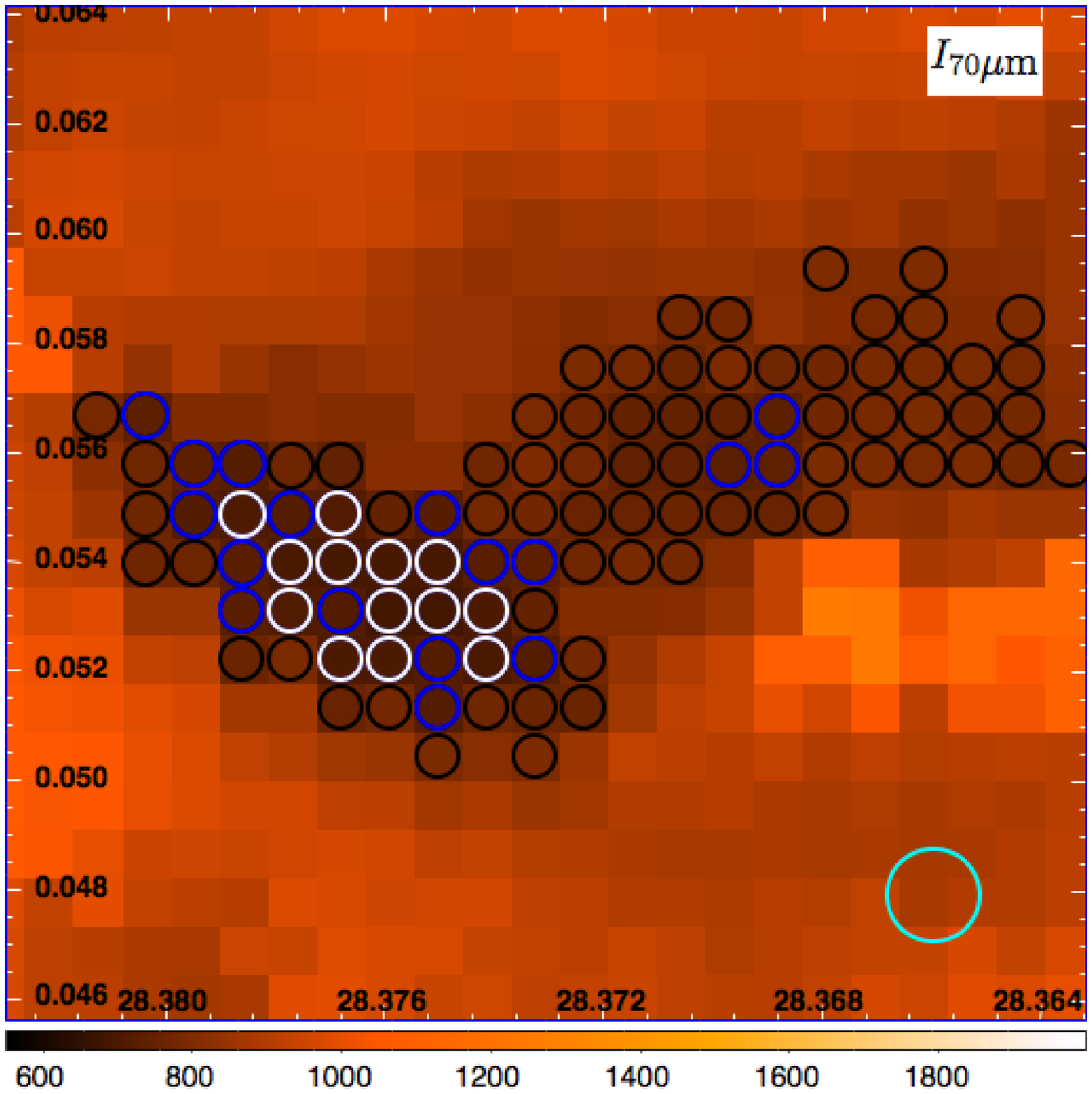} \\
\hspace{-0.1in} \includegraphics[width=2.15in]{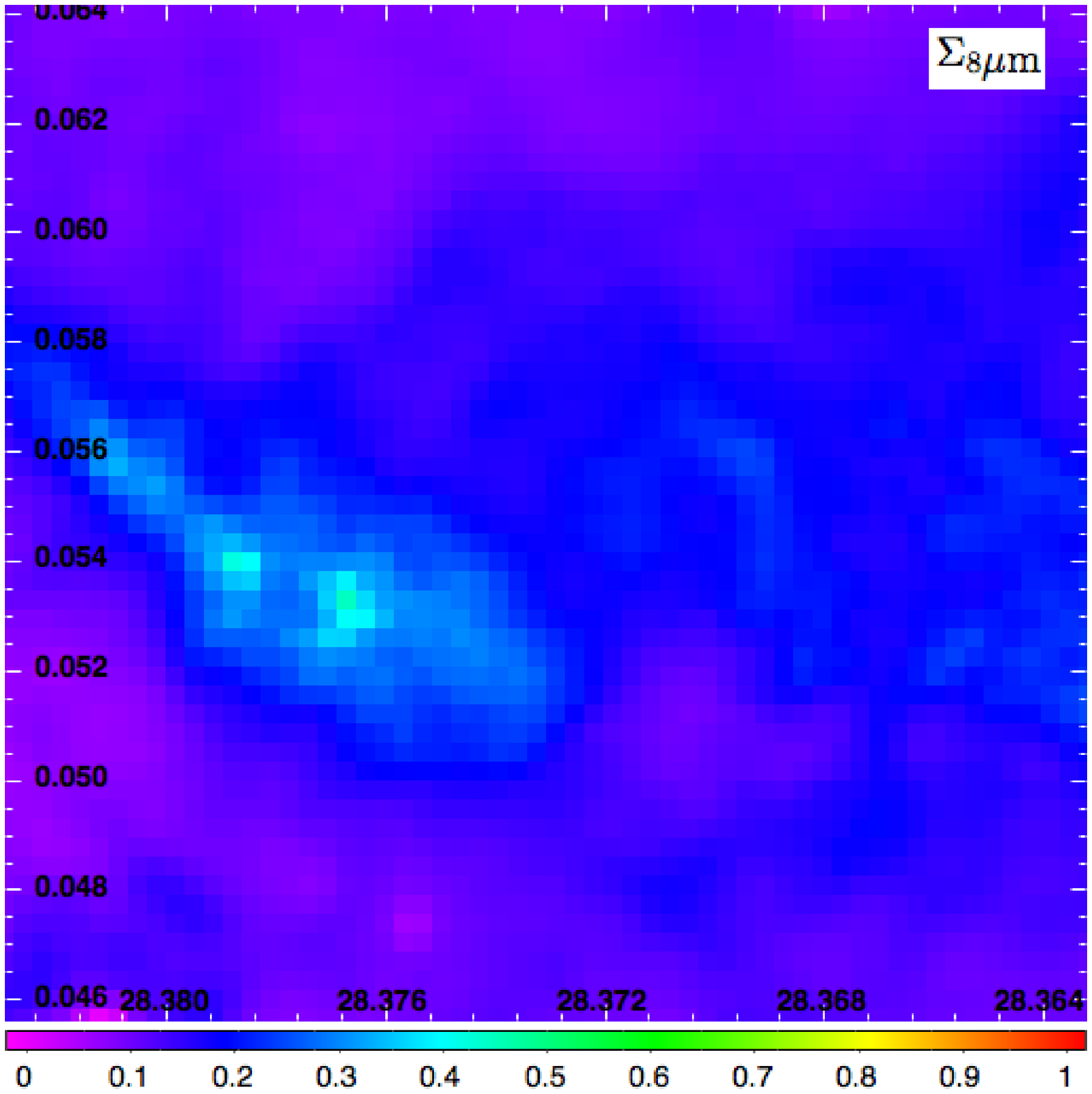} &\hspace{-0.1in} \includegraphics[width=2.15in]{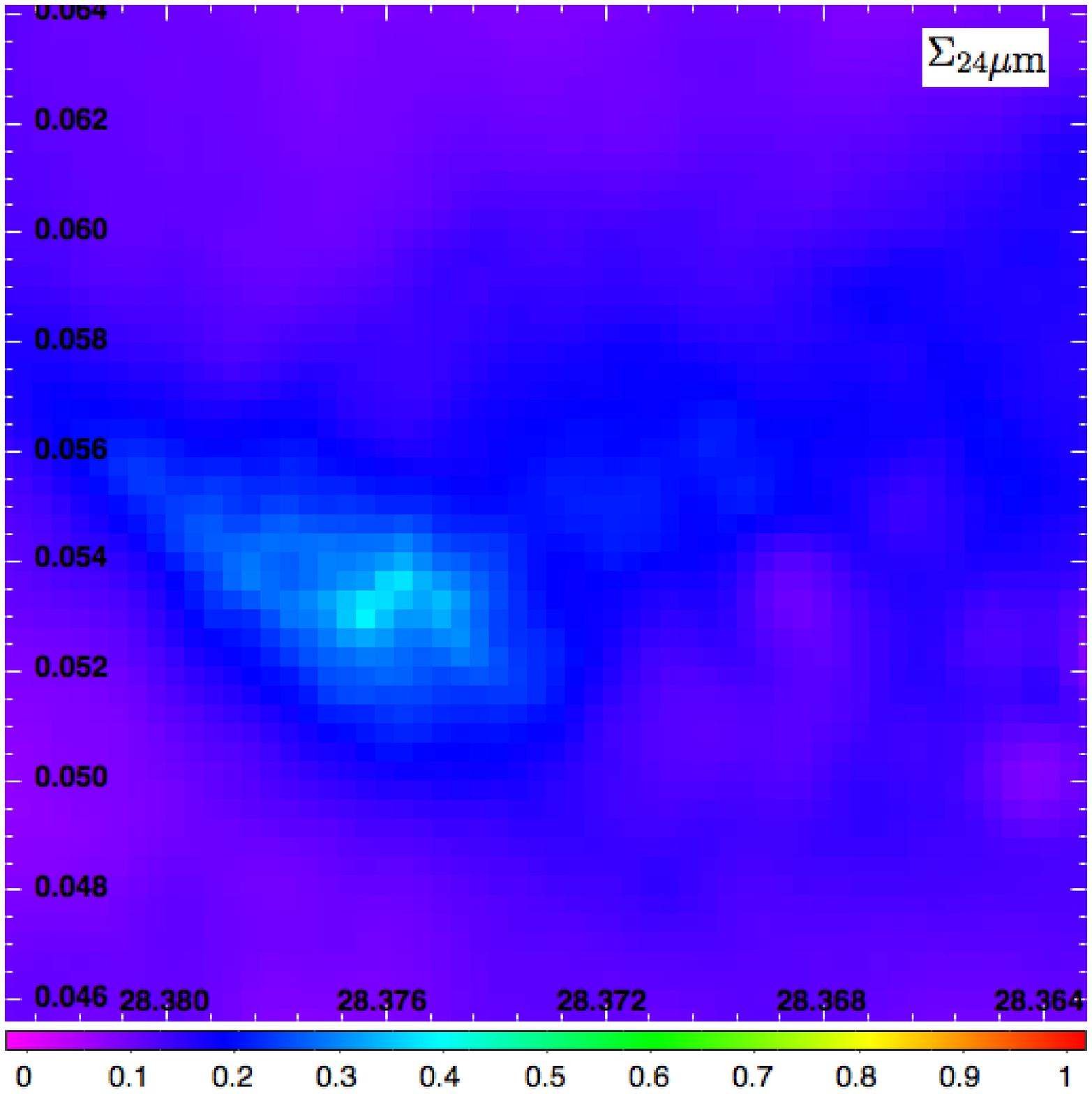} &\hspace{-0.1in} \includegraphics[width=2.15in]{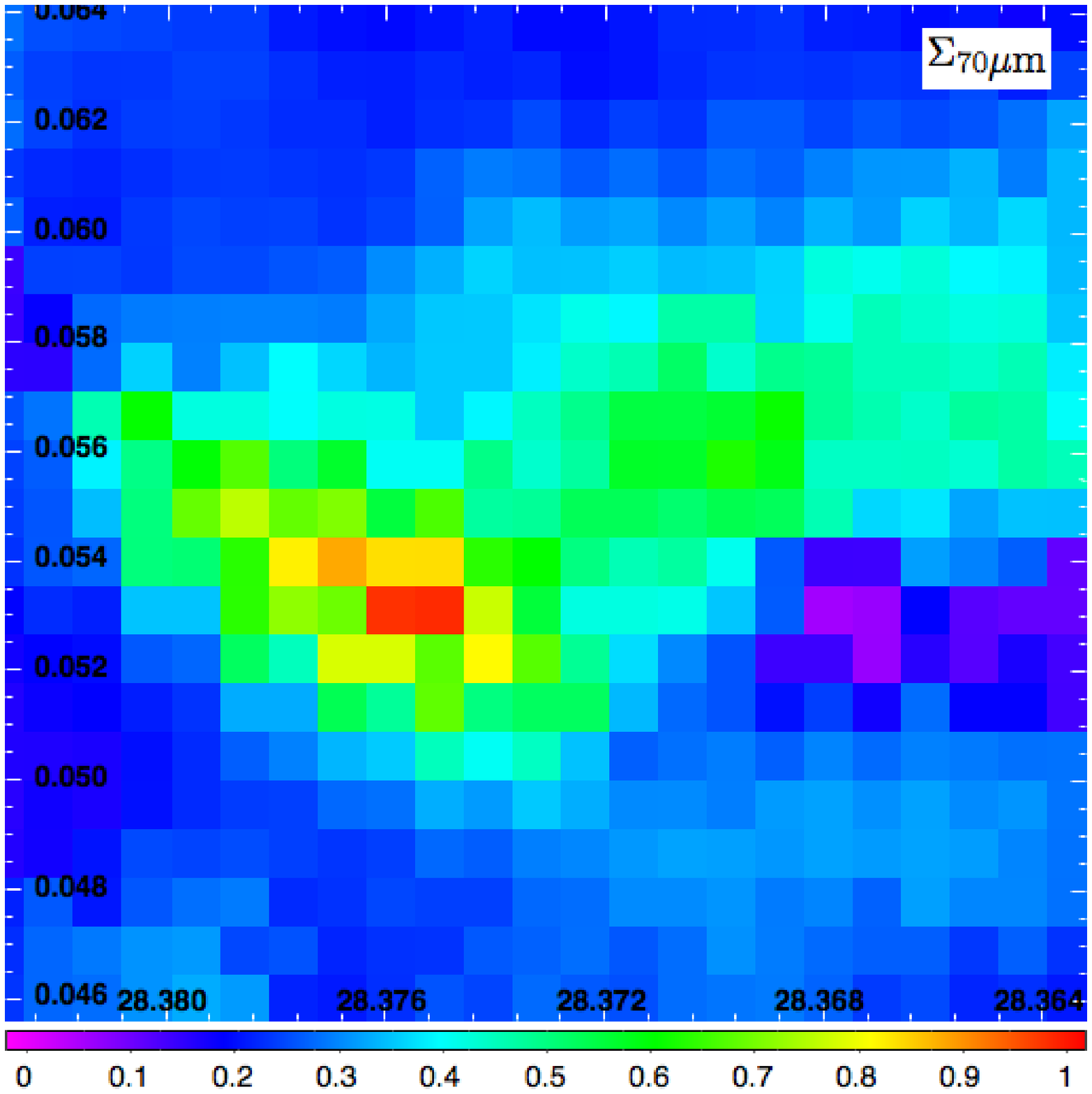} \\
\hspace{-0.1in} \includegraphics[width=2.15in]{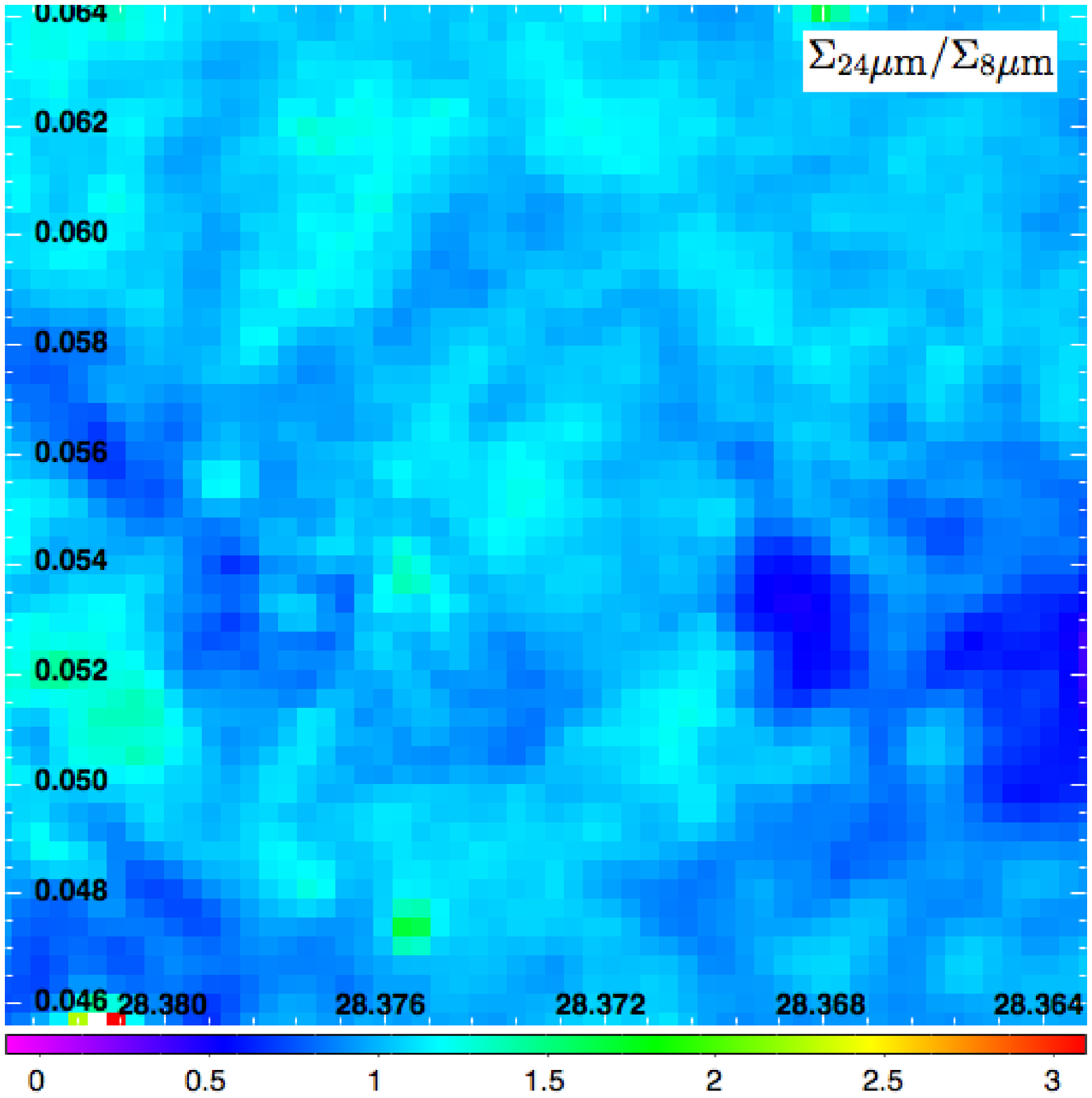} &\hspace{-0.1in} \includegraphics[width=2.15in]{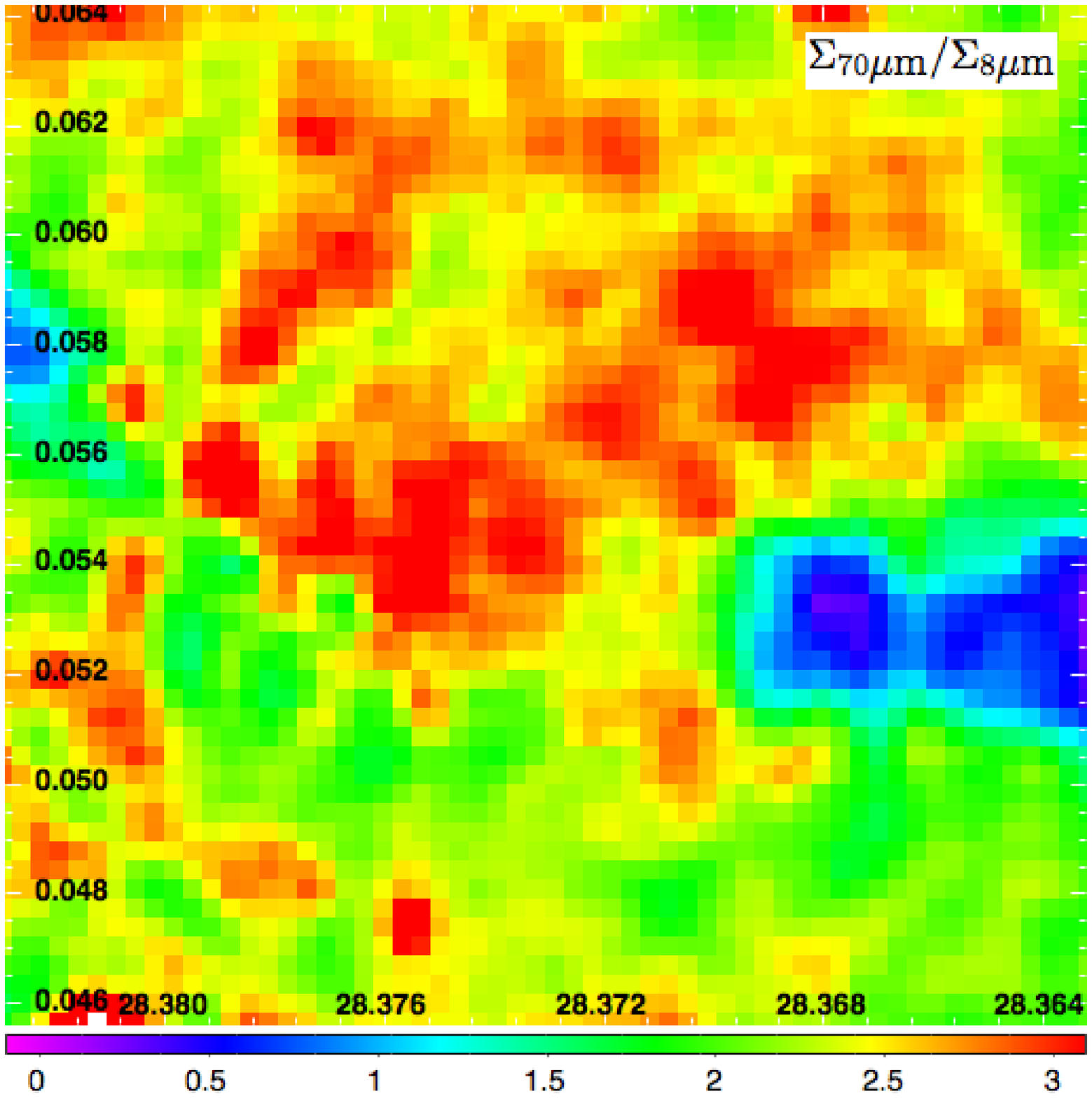} &\hspace{-0.1in} \includegraphics[width=2.15in]{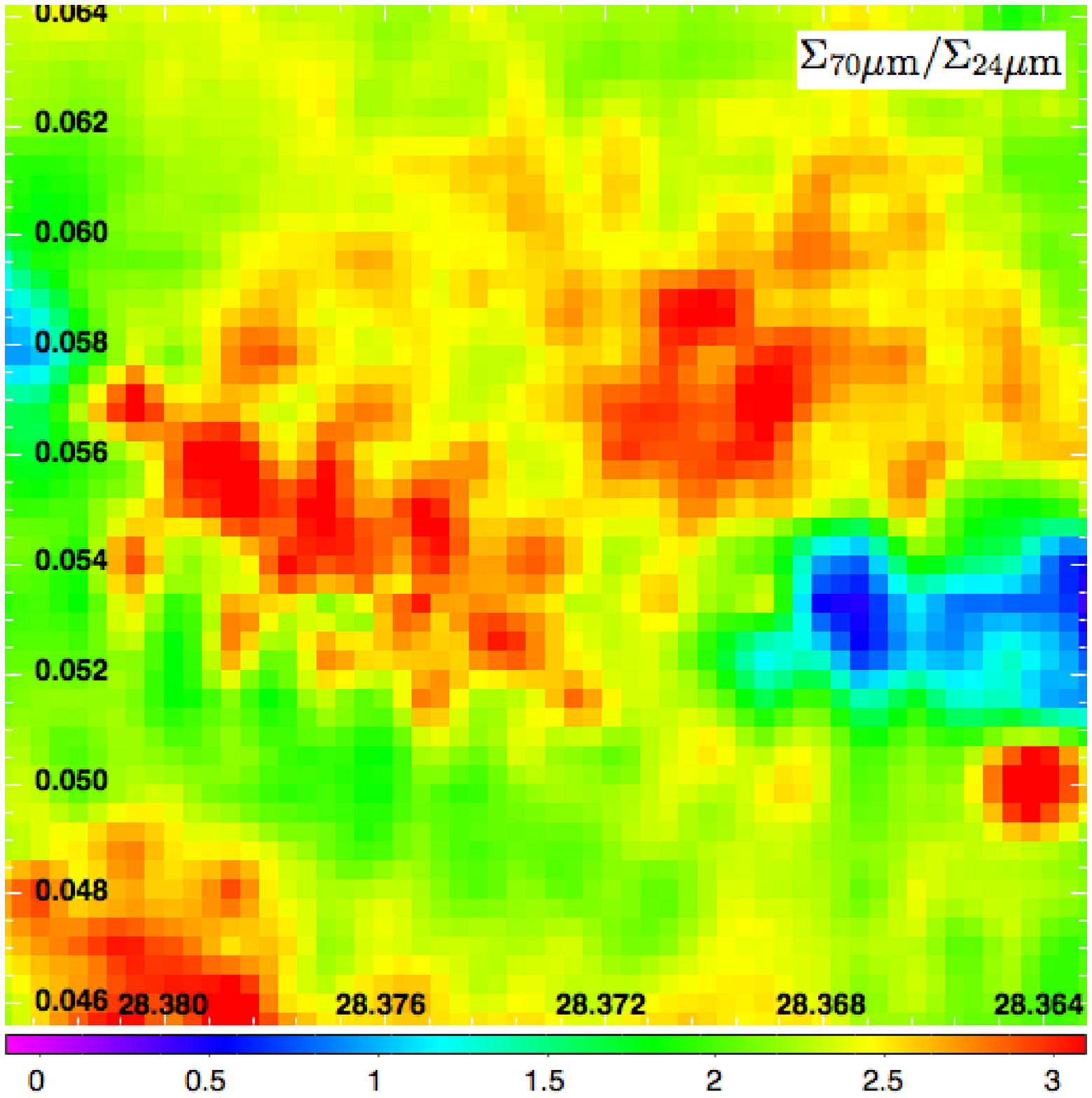} \\
\end{array}$
\end{center}
\caption{\footnotesize Same with Figure~\ref{fig:c1} but for C11.}
\label{fig:c11}
\end{figure}

The relative values of $\Sigma$ derived in non-saturated regions at
different wavelengths yield information about the shape of the dust
extinction law. This law is expected to vary as grains undergo growth
via coagulation and increasing deposition of volatiles, such as water,
methanol and CO, to form ice mantles (see Fig.~\ref{fig:wholecloud}d).
The $\Sigma$ maps of Figs.~\ref{fig:c1}, \ref{fig:c4} and
\ref{fig:c11} are derived assuming the moderately-coagulated thick ice
mantle model of OH94, so deviations in the $\Sigma$ ratios,
e.g. $\Sigma_{70\micron}/\Sigma_{8\micron}$, from unity tell us about
deviations of the actual dust extinction law from the OH94 model.

In the bottom rows of Figs.~\ref{fig:c1}, \ref{fig:c4} and
\ref{fig:c11} we present maps of
$\Sigma_{24\micron}/\Sigma_{8\micron}$,
$\Sigma_{70\micron}/\Sigma_{8\micron}$ and
$\Sigma_{70\micron}/\Sigma_{24\micron}$. Some of the variation in
these ratio maps is due to the different saturation levels, i.e. the
70\micron\ map is able to probe to $\Sigma\simeq1\:{\rm{g\:cm^{-2}}}$,
while the other maps saturate at $\sim0.4\:{\rm{g\:cm^{-2}}}$ (and
tend to have ratio maps closer to unity). Thus we focus on variations
present in non-saturated regions, especially considering intervals of
$\pm0.05\:{\rm{g\:cm^{-2}}}$ centered on $\Sigma_{8\micron} =
0.1,0.2,0.3\:{\rm{g\:cm^{-2}}}$. For pixels in these ranges, we
evaluate the mean values of $\Sigma_{24\micron}/\Sigma_{8\micron}$
and $\Sigma_{70\micron}/\Sigma_{8\micron}$. These mean ratios can be
reconciled to unity by assuming different values of
$\kappa_{24\micron}$ and $\kappa_{70\micron}$ relative to
$\kappa_{8\micron}$, and these are shown in Fig.~\ref{fig:wholecloud}d
for each of the three cores. 

We also derived $\Sigma$ maps in IRAC bands 1, 2 and 3 for each of the
three regions, following the same methods described above and in BT12,
and used their ratio maps compared to $\Sigma_{8\micron}$ to explore
opacity variations down to 3.5\micron\ (Fig.~\ref{fig:wholecloud}d).
Our fiducial maps use the total stars plus dust SED of the Galactic
background. If the spectrum of only the dust component is used (see
Fig.~\ref{fig:wholecloud}b), then the effective opacity would change by
at most 10\%.

The relative opacity values from the MIR to FIR, generally follow the
thick ice mantle models of OH94. In particular the ratio of
$\kappa_{70\micron}/\kappa_{8\micron}$ and, to a lesser extent, that
of $\kappa_{24\micron}/\kappa_{8\micron}$, favor such models.  At
shorter wavelengths, the thick and thin ice mantle models are more
similar, so there is less discriminatory power, although there is a
hint that the 3.5\micron\ IRAC band is picking up growth of the
3\micron\ water ice feature.

There are hints of a systematic increase in
$\kappa_{70\micron}/\kappa_{8\micron}$ with increasing $\Sigma$,
especially in the regions around cores C1 and C4, which would be
consistent with basic expectations of grain evolution (OH94). However,
a larger sample of regions is needed to confirm this trend. 

It is possible that systematic differences seen from core region to
core region (e.g., the C1 region shows relatively lower 6 and
70\micron\ opacities) could reflect real evolutionary differences
between the regions. However, systematic temperature variations, if
extending $\gtrsim15\:$K \citep{tan2013b} could affect FIREX maps at
70\micron\ and produce similar effects.



The opacity features in the thick ice mantle models around 3, 6, 20
and 50\micron\ are mostly caused by H$_2$O ice, but CH$_3$OH ice makes
an $\sim1/3$ contribution to the 6\micron\ feature
\citep{hudgins1993}. It is possible that astrochemical variations in
the abundance of CH$_3$OH could be contributing to the observed
dispersion of the 6\micron\ opacity, and potentially also affecting the
8\micron\ opacity, which sets the normalization of the curves and data
shown in Fig.~\ref{fig:wholecloud}d. Again a larger sample of regions
and a search for potential correlations with other astrochemical
indicators is needed to assess these possibilities.


\acknowledgements We thank Michael Butler, Sean Carey, Bruce Draine,
Elizabeth Lada and the referee for helpful discussions and
comments. We acknowledge grants from Florida Space Inst. and NASA
(ADAP10-0110).

\end{document}